\documentclass[aps,prb,amsmath,amssymb,longbibliography,lengthcheck,showpacs,superscriptaddress,floatfix]{revtex4-1}
\usepackage{graphicx}
\usepackage{subfigure}
\usepackage{color}
\newcommand{\deri}[2]{{\displaystyle \frac{\partial #1 }{\partial #2 }}}
\newcommand{\derri}[2]{{\displaystyle \frac{\partial^2 #1 }{\partial #2 }}}
\begin{document}
\title{How are vibrational excitations and thermal conductivity related to elastic heterogeneities in disordered solids?}
\author{Hideyuki Mizuno}
\email{hideyuki.mizuno@phys.c.u-tokyo.ac.jp}
\altaffiliation{Current address: Graduate School of Arts and Sciences, The University of Tokyo, Tokyo 153-8902, Japan}
\affiliation{Univ. Grenoble Alpes, LIPHY, F-38000 Grenoble, France}
\affiliation{CNRS, LIPHY, F-38000 Grenoble, France}
\author{Stefano Mossa}
\email{stefano.mossa@cea.fr}
\affiliation{Univ. Grenoble Alpes, INAC-SYMMES, F-38000 Grenoble, France}
\affiliation{CNRS, INAC-SYMMES, F-38000 Grenoble, France}
\affiliation{CEA, INAC-SYMMES, F-38000 Grenoble, France}
\author{Jean-Louis Barrat}
\email{jean-louis.barrat@ujf-grenoble.fr}
\affiliation{Univ. Grenoble Alpes, LIPHY, F-38000 Grenoble, France}
\affiliation{CNRS, LIPHY, F-38000 Grenoble, France}
\affiliation{Institut Laue-Langevin - 6 rue Jules Horowitz, BP 156, 38042 Grenoble, France}
\date{\today}
\begin{abstract}
In crystals, molecules thermally vibrate around the periodic lattice sites. Vibrational motions are well understood in terms of phonons, which carry heat and control heat transport. The situation is notably different in disordered solids, where vibrational excitations are not phonons and can be even localized. Recent numerical work has established the concept of elastic heterogeneity: Disordered solids show inhomogeneous local mechanical response. Clearly, the heterogeneous nature of elastic properties strongly influences vibrational and thermal properties, and it is expected to be the origin of anomalous features, including boson peak, vibrational localization, and temperature dependence of thermal conductivity. These are all crucial long-standing problems in material physics, which we address in the present work. We have considered a toy model able to stabilize different states of matter, by introducing an increasing amount of size disorder. The phase diagram generated by Molecular Dynamics simulation encompasses the perfect crystalline state with spatially homogeneous elastic moduli distribution, multiple defective phases with increasing moduli heterogeneities, and eventually a series of amorphous states. We have established clear correlations among heterogeneous local mechanical response, vibrational states, and thermal conductivity. We provide evidence that elastic heterogeneity controls both vibrational and thermal properties, and is a key concept to understand the anomalous puzzling features of disordered solids.
\end{abstract}
\maketitle
\tableofcontents
\section{Introduction}
\label{sect:introduction}
In crystalline materials, molecules are located at the periodic lattice sites, and their vibrational motions are well understood in terms of quantized plane waves, the phonons~\cite{kettel,Ashcroft}. At low frequencies ($\omega$), vibrations are described as {\em acoustic} plane waves, whose vibrational density of states (vDOS) conforms to the Debye model, $g_D(\omega)\propto\omega^2$, which agrees with experimental results for crystals~\cite{kettel,Ashcroft}. In contrast, disordered solids feature vibrational properties anomalous compared to those of the corresponding crystals. (Here, disordered solids include not only topologically amorphous materials as structural glasses~\cite{lowtem}, but also disordered crystals~\cite{Kaya_2010,Green_2011}, which show periodic lattice structures but in the presence of disordered inter-particle potentials, like colloidal crystals with size disorder.) Among these anomalies, the origin of an excess in the low-$\omega$ spectrum of the excitations, the boson peak (BP)~\cite{buchenau_1984,Malinovsky_1991}, is still an open issue. More precisely, $g(\omega)$ shows an excess over the Debye prediction for the corresponding crystal value, around a frequency $\omega=\Omega^\text{BP} \sim 1$~THz. At $\Omega^\text{BP}$, vibrational excitations can even be localized~\cite{mazzacurati_1996,Schober_2004}, and in general cannot be described as plane waves.

Interestingly, {\em acoustic-like} excitations have been observed in disordered solids by experimental techniques, including light~\cite{Shapiro_1966}, and (inelastic) X-rays~\cite{Sette_1998} and Neutrons~\cite{Bove_2005} scattering. Numerical methods like Molecular Dynamics (MD) simulations~\cite{Rahman_1976} have also provided clear evidences in this direction. In the case of crystals, acoustic excitations are exact normal modes of the system, and an acoustic plane wave excites one normal mode only. In contrast, an acoustic-like vibrational excitation in disordered solids is a superposition of several different normal modes, with different vibrational frequencies~\cite{taraskin_2000,matsuoka_2012}. Such the mode attenuates rather rapidly~\cite{matsuoka_2012} compared to vibrations in crystals. It has been reported that the Ioffe-Regel frequency, $\Omega^\text{IR}$, which corresponds to an upper bound for the frequency of propagation of true plane waves~\cite{Ioffe_1960}, is located around the BP frequency, $\Omega^\text{IR} \sim \Omega^\text{BP}$~\cite{ruffle_2006,shintani_2008,beltukov_2016}. More interestingly, strong scattering and breakdown of the Debye-continuum approximation have been observed around the same frequency~\cite{monaco2_2009}. Connections between the BP and anomalous acoustic excitations are not obvious, but the above observations indicate that they must be strongly correlated.

Anomalies in vibrational properties obviously reflect on thermal behaviour~\cite{lowtem}, including heat capacity and thermal conductivity at low temperature, $T$. The low-$T$ heat capacity, $C(T)$, can be directly obtained from the $g(\omega)$ in the harmonic approximation~\cite{kettel,Ashcroft} (see Eq.~(\ref{spheat})). From the Debye prediction, $g_D(\omega)$, one obtains $C_D(T)\propto T^3$, which captures well the low-$T$ heat capacity of crystals~\cite{kettel,Ashcroft}. In contrast, disordered solids show values higher than the Debye prediction, which directly originates from the excess vibrational modes in the BP frequency range~\cite{lowtem,Zeller_1971,Schober_1996}.

Thermal conductivity, $\kappa$, is also very different in crystals and disordered solids. In crystals, phonons carry heat and play the most important role in thermal conduction~\cite{kettel,Ashcroft}. Therefore, although it is in principle necessary to correctly take into account phonon-phonon interactions at non-zero temperatures (anharmonic effects), one can very precisely analyse thermal conductivity in terms of the Boltzmann transport equation for phonons~\cite{McGaughey_2004,McGaughey,Broido_2007,Turney2_2009}. In disordered solids, the nature of heat carriers is still matter of debate, but acoustic-like modes are naturally expected to play an important role. In this case, a strong damping of acoustic-like excitations with $\omega >\Omega^\text{BP} \sim \Omega^\text{IR}$ leads to an important reduction of $\kappa$~\cite{lowtem,Zeller_1971,Cahill_1988,Cahill_1992}. Remarkably, disordered solids generally show similar temperature dependence of thermal conductivity, irrespective to the details of their chemical structure. More precisely, $\kappa$ increases as $\kappa \sim T^2$ at low-$T$, and exhibits a plateau around $T\sim 10$~K~\cite{lowtem,Zeller_1971,Cahill_1988,Cahill_1992}. A theoretical calculation for $\kappa$ of disordered solids has been proposed, where heat currents are carried by non-propagating, delocalized, normal modes, called diffusons~\cite{Allen_1993}. This theory is able to reproduce the $T$-dependence of thermal conductivity in the glass phase~\cite{Feldman_1993}.

Clarifying the above issues is tantamount with seeking an answer to the question: What is the origin of vibrational and thermal anomalies in disordered solids? This issue has been targeted by several theoretical developments. These include, among others, the soft-potential model~\cite{Karpov_1983}, the Mode Coupling Theory~\cite{Gotze_2000}, crossover from minima-dominated to saddle-point-dominated phases~\cite{Grigera_2003}, vibrational instability of quasi-localized modes~\cite{Gurevich_2003}, transformation of van Hove singularities~\cite{Schirmacher_1998,Taraskin_2001}, piling-up of acoustic states close to the boundary of the pseudo-Brillouin zone~\cite{Chumakov_2011,Chumakov_2014}, weak connectivities of particles due to the vicinity of the jamming transition point~\cite{Wyart_2005,Wyart_2006}. 

In addition, the concept of {\em elastic heterogeneity} has been proposed~\cite{Duval_1998}: Disordered solids exhibit spatial heterogeneities of elastic moduli. This is a specific feature, absent in ordered crystals where the mechanical response to perturbations is homogeneous at all length scales~\cite{Wagner_2011,Hufnagel_2015}. Recent simulation works~\cite{yoshimoto_2004,tsamados_2009,makke_2011,Mizuno_2013} have addressed a direct measure of {\em local} elastic moduli, and have well established this concept. (Note that local measurements of elastic properties can be quite easily implemented in numerical simulations~\cite{Mizuno_2013}, whereas analogous experimental measurements are rather difficult~\cite{Wagner_2011}.) The study of Ref.~\cite{tsamados_2009} showed that local moduli spatially fluctuate at mesoscopic length-scales, $\xi_{\text{eh}} \sim 10$ to $15 \sigma$, with $\sigma$ the typical atomic diameter. Also, Refs.~\cite{DiDonna_2005,Maloney_2006} showed that the spatial heterogeneities in elastic properties generate non-affine deformations, which add to the affine contributions and are of comparable magnitude~\cite{tanguy_2002}. During the non-affine deformations, particles have been shown to undergo correlated displacements, with a mesoscopic correlation length, $\xi_\text{na} \sim 20$ to $30 \sigma$~\cite{tanguy_2002,leonforte_2005}, which is of the same order of magnitude as $\xi_{\text{eh}}$.

It is natural to expect that elastic heterogeneities must contribute in turning phonons to more complex vibrational excitations, therefore scattering acoustic plane-waves and reducing thermal conductivity. Remarkably, it was reported that the wavelength $\Lambda$ of acoustic waves corresponding to $\Omega^\text{BP}$, is close to the mesoscopic length-scale $\xi_{\text{na}}$~\cite{leonforte_2005,leonforte_2011}, i.e., $\Lambda \sim \xi_{\text{na}} \sim \xi_{\text{eh}}$. The breakdown of continuum elasticity~\cite{wittmer_2002,tanguy_2002} and Debye-approximation~\cite{monaco_2009} for acoustic plane-waves, and the onset of the strong scattering regime~\cite{Baldi_2013} have been also found to take place at similar length-scales as $\xi_{\text{eh}}$ and $\xi_\text{na}$. Also, strong correlations between local moduli and vibrational modes have been detected: Localization of vibrational excitations tends to appear in soft regions, characterized by elastic constants significantly lower than the macroscopic values~\cite{tanguy_2010,Derlet_2012}. A theoretical approach based on the concept of spatially fluctuating elastic moduli~\cite{schirmacher_2006,schirmacher_2007,Schirmacher_2015,Schirmacher2_2015} has been able to reproduce both the BP feature and the $T$-dependence of thermal conductivity. All this work therefore supports the hypothesis that elastic heterogeneities control both vibrational and thermal anomalies. 

In recent works~\cite{Mizuno2_2013,Mizuno_2014} we have addressed this point, by systematically modulating the extent of the heterogeneous elastic response. We have provided evidence of direct correlations with vibrational states features and thermal conductivity, determined by completely independent calculations without any adjustable parameter. Our approach was based on Molecular Dynamics (MD) simulations of a toy model, which allowed us to generate states of matter ranging from the perfect crystal state to defective crystal phases, and eventually, amorphous states, by introducing an increasing amount of disorder in particles size. Next, we: {\em i)} characterized the changes of elastic moduli heterogeneities in the different phases; {\em ii)} independently studied the consequent modifications of vibrational excitations, both in terms of eigenvalues and eigenvectors of the Hessian matrix and spectroscopic parameters extracted from dynamical structure factors; and {\em iii)} monitored the associated changes in the $T$-dependence of thermal conductivity. 

Here, we present significantly  more extended data sets, explore in details the nature of vibrational excitations and their correlation with different local elastic constants in various regions of the spectrum, clarify the effect of anharmonic couplings, and offer a general perspective on our work. The paper is organized as follows. In Section~\ref{sect:methods}, we describe our numerical model, and give details about the method used to measure the local elastic constants. In Section~\ref{sect:elastic heterogeneities}, we present a discussion of our results on elastic heterogeneities. We also attempt to correlate heterogeneities of local moduli to those present in more familiar local structural quantities. In Sections~\ref{sec.vs} and~\ref{sec.thc}, we present the results on vibrational states (vDOS, participation ratios, life-times) and thermal conductivity, respectively, and detail the correlations between the elastic heterogeneities on one side, and vibrational states and thermal conductivity, on the other. Finally, in Section~\ref{sect:conclusions}, we summarize our results and draw general conclusions on our work.
\begin{figure}[t]
\centering
\includegraphics[width=0.48\textwidth]{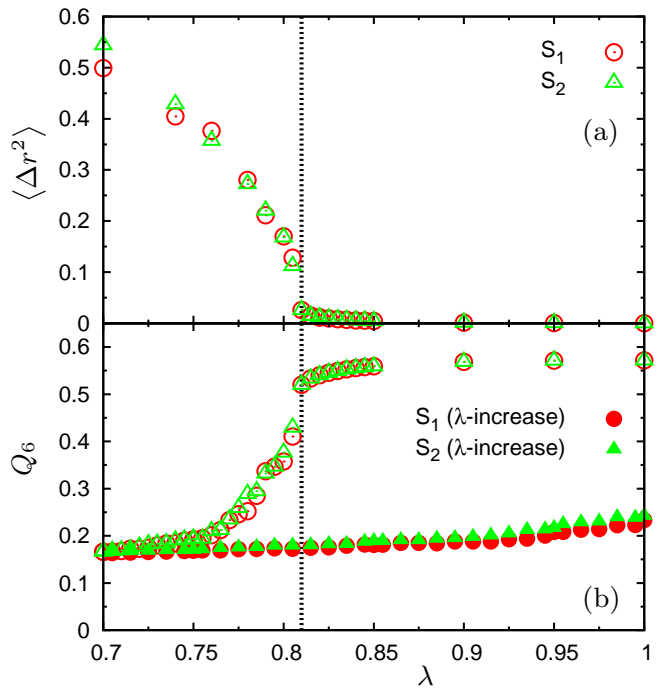}
\caption{
(a) Mean-squared displacement $\left< \Delta r^2\right>$ and (b) $Q_6$, versus the disorder parameter $\lambda$, at $T=10^{-2}$. The system is initialized in the perfect FCC-crystal state $\lambda =1$, where $\left< \Delta r^2\right>\simeq 0$ and $Q_6 \simeq 0.57$. Next, $\lambda$ is decreased from $1$ to $0.7$ in the fully developed amorphous state, as discussed in the text. Data for two independent instances of the size disorder are shown by open circles and triangles. Both samples undergo the amorphisation transition at $\lambda = \lambda^\ast \simeq 0.81$, indicated by the vertical line. We also show in (b) by closed circles data for $Q_6$ obtained by following the reverse path,  increasing $\lambda$ from $0.7$ to $1$. In this case, $Q_6$ shows no significant changes for $\lambda = \lambda^\ast$, keeping the value pertaining to the amorphous state. This hysteresis effect is discussed in the text. 
}
\label{transition1}
\end{figure}
\section{Numerical Methods}
\label{sect:methods}
\subsection{Model and simulations details}
{\bf Soft spheres.} We have considered a soft-sphere model~\cite{bernu_1987} in a (3D) cubic box of linear size $L$, with periodic boundary conditions in all directions. Particles $i$ and $j$ interact through a soft-sphere potential,
\begin{equation}
v^{ij} = \epsilon \left( \frac{\sigma^{ij}}{r^{ij}} \right)^{12},
\end{equation}
where $\sigma^{ij} = (\sigma^{i} + \sigma^{j})/2$, $\sigma^{i}$ and $\sigma^{j}$ are the diameters of the particles, and $r^{ij}$ is the mutual distance. The potential $v^{ij}$ is cut-off and shifted to zero at $r_c^{ij} = 2.5 \sigma^{ij}$. Our reference state is the one-component perfect face-centred-cubic (FCC) crystal, where the particle diameter and mass are $\sigma$ and $m$ for all particles. Throughout this study, we use $\sigma$, $\epsilon/k_B$ ($k_B$ is the Boltzmann constant), and $\tau =(m\sigma^2/\epsilon)^{1/2}$ as units of length, temperature, and time, respectively, i.e., we set $\sigma=\epsilon=\tau=1$.

We have fixed the number density $\hat{\rho}=N/V=N/L^3=1.015$ ($N$ is number of particles, and $V$ is the system volume), and the length of the unit cell of the FCC crystal is $a = 1.58$. Most of the  simulations were performed with $N=4000$ particles, in boxes of linear size $L=10 a=15.8$. Larger systems, with $L$ ranging from $L=12a$ ($N=6912$) to $30a$ ($N=108000$), were also used for the calculations of the vibrational states (see Sec.~\ref{sec.vs}). The FCC crystal was equilibrated at temperature $T=10^{-2}$ in the $(NVT)$ ensemble, by using a Berendsen thermostat~\cite{Berendsen_1984}. Although we set the number density $\hat{\rho}$ and the temperature $T$ independently, the thermodynamic state of the present system depends on a single parameter, $\Gamma = \hat{\rho}/T^{1/4}$, due to the scaling properties of inverse-power-law potentials~\cite{bernu_1987}. $\Gamma =3.21$ in the present case. For a one component soft-sphere system, melting and glass transition temperatures are $T_m \simeq 0.6$ ($\Gamma_m \simeq 1.15$) and $T_g \simeq 0.2$ ($\Gamma_g \simeq 1.5$), respectively~\cite{bernu_1987}. All simulations have been performed by using the MD code LAMMPS~\cite{Plimpton_1995,lammps}.

{\bf Size disorder.} Starting from the reference perfect crystal state, we introduce  disorder in particle size, as described in Ref.~\cite{bocquet_1992}. We randomly select $N/2$ particles which are assigned to species $1$ with size $\sigma_1$, the remaining pertaining to species $2$ ($\sigma_2$), therefore designing an initial equimolar binary mixture. In an approximate one-component description, an effective diameter can be defined as $\sigma_{\text{eff}}^3 = \sum_{\alpha,\beta=1,2} x_\alpha x_\beta \sigma_{\alpha \beta}^3$, where $\sigma_{\alpha \beta} = (\sigma_{\alpha} + \sigma_{\beta})/2$ and $x_\alpha=x_\beta=0.5$ are the respective molarities~\cite{bernu_1987}. The coupling parameter $\Gamma$ is therefore replaced by $\Gamma_\text{eff}=(\hat{\rho}/T^{1/4})\sigma^3_\text{eff}$. Next, $\sigma_1$ is gradually reduced below the initial value $\sigma=1$, while $\sigma_2$ is increased above $1$, by keeping constant both the effective diameter $\sigma_\text{eff} \equiv 1$ and the coupling parameter $\Gamma_\text{eff} \equiv 3.21$. The extent of the disorder is therefore encoded in the {\em disorder parameter}, $\lambda=\sigma_1/\sigma_2 \le 1$, which directly provides the values of $\sigma_1$ and $\sigma_2$. We started with the ideal crystal case, $\lambda=1$, and gradually decreased $\lambda$ by a series of small steps, $\Delta \lambda=10^{-4}$, encompassing the range $\lambda\in [0.7:1]$. The system was re-equilibrated at $T=10^{-2}$ after each step before production runs. We note that the total volume fraction $\phi$ varies only mildly ($\phi=53$ to $56 \%$) during the entire process (see Fig.~\ref{ehorder1}(b)).

{\bf The amorphisation transition.} As the disorder parameter $\lambda$ is decreased, therefore introducing an increasing size disorder, the system undergoes a structural transition into an amorphous state at $\lambda=\lambda^\ast$, as first observed in Ref.~\cite{bocquet_1992}. Note that, although Ref.~\cite{bocquet_1992} considered a 2D system, the result is very similar for our 3D case. We determined the transition point $\lambda^\ast\simeq 0.81$ by monitoring both the mean-squared displacement $\left< \Delta r^2 \right> = (1/N) \sum_{j=1}^N( \left< \mathbf{r}^j \right> - \mathbf{r}^{j}_{0})^2$ and the bond order parameter $Q_6$~\cite{Steinhardt_1983, Lechner_2008}. Here, $\left< \right>$ denotes the time average (ensemble average), $\mathbf{r}^j$ is the instantaneous position of particle $j$, and $\mathbf{r}^{j}_{0}$ is the reference FCC lattice site. 

In Fig.~\ref{transition1} we show the $\lambda$-dependence of $\left< \Delta r^2 \right>$ and $Q_6$, by open symbols. For the perfect crystal, $\lambda=1$, we have $\left< \Delta r^2 \right> = 0$ and $Q_6 \simeq 0.57$. As $\lambda$ decreases, both quantities show discontinuous jumps at the transition point, $\lambda^\ast$. We have additionally monitored the order parameter $\left< \left| \rho_{\mathbf{G}} \right| \right>=\left< \left| (1/N) \sum_{j=1}^N \exp(i \mathbf{G}\cdot \mathbf{r}^j ) \right| \right>$, with $\mathbf{G}=(2\pi/a,2\pi/a,-2\pi/a)$, which also shows a discontinuity at $\lambda^\ast$~\cite{Mizuno2_2013}. The value of $\lambda^\ast$ does not depend on the initial repartition of the two species on the lattice, as we demonstrate in Fig.~\ref{transition1} where we show analogous results for two independent instances of the disorder. 

For $\lambda^\ast \le \lambda\le 1$, particles are localized very close to the initial lattice sites, notwithstanding the presence of size disorder. The system is therefore in a chemically-disordered crystalline state~\cite{Kaya_2010,Green_2011}, characterized by well-defined Bragg peaks. In contrast, for $0.7 \le \lambda< \lambda^\ast$, the system cannot keep the lattice structure any longer, and falls in an amorphous arrested state, with complete loss of translational invariance. As discussed in Ref.~\cite{bocquet_1992}, the transition is first-order-like. Indeed, the first derivative of the free energy with respect to $\lambda$, $\partial F/\partial \lambda$, exhibits a discontinuous change at $\lambda^\ast$~\cite{bocquet_1992}, which is a behaviour typical of a genuine first-order phase transition. The parameter $\lambda$, however, is {\em not} a true thermodynamic variable, and therefore the transition cannot be strictly considered as such in a genuine thermodynamic sense (see Ref.~\cite{bocquet_1992} for details).

{\bf Hysteresis.} It is interesting to reversely increase $\lambda$, searching for hysteresis effects. We show our result in Fig.~\ref{transition1}(b) (filled symbols). Interestingly, $Q_6$ shows no significant changes in the entire $\lambda$-range. This means that, at the investigated low $T$, the system is trapped in the amorphous state and cannot overcome the energy barrier leading to the crystalline minimum, at least on our simulation time scale. Indeed, we have also confirmed that at $T=10^{-1}$, the system partially recovers the lattice structure, at $\lambda\simeq 0.95$, but still cannot return to the perfect crystal state. This is at variance with Ref.~\cite{bocquet_1992}, where a reinitialization to the perfect lattice structure upon increasing $\lambda$ was observed. This difference can be explained by observing that in the 2D case for small system with $N=108$~\cite{bocquet_1992}, the energy barrier separating the amorphous and crystalline states can be expected to be much smaller than that of the present 3D case with $N=4000$. Our results are also consistent with those of Refs.~\cite{Hamanaka_2006,Hamanaka_2007}, where a larger 2D system with $N=1000$ was studied varying both $T$ and $\lambda$, and poly-crystalline domains separated by amorphous boundaries were reported.
\subsection{Measuring the local elastic moduli} 
\label{sec.measure}
Disordered solids, including glasses and complex crystals, exhibit inhomogeneous and scale-dependent spatial distributions of local elastic moduli. These can be measured following different methods~\cite{yoshimoto_2004,tsamados_2009,makke_2011,Mizuno_2013}. In the present study, we employ the equilibrium fluctuation formulae, which can be used to calculate both global~\cite{squire_1969,ray_1984,barrat_1988,lutsko_1989,Wittmer_2013} and local~\cite{lutsko_1988,yoshimoto_2004,Mizuno_2013} moduli. In Ref.~\cite{Mizuno_2013}, we referred to this method as the {\em fully-local} approach, which we summarize below.

{\bf The local modulus tensor.} The local elastic response at a coarse-graining length scale $w$ can be determined by partitioning the simulation box into $20^3$ cubic domains, identified by the index $m$, of linear size $w=2a=3.16$. A domain has a volume $w^3=8a^3$, which is $8$ times that of the unit cell of the FCC crystal, and includes about $30$ particles. The local modulus tensor ${C}^m_{\alpha \beta \gamma \delta}$ ($\alpha,\beta,\gamma,\delta=x,y,z$) is defined as the derivative of the local stress ${\sigma}^m_{\alpha \beta}$ with respect to the local (linear) strain ${\epsilon}^m_{\gamma \delta}$, and can be expressed as:
\begin{equation}
\begin{aligned}
C_{\alpha \beta \gamma \delta}^m &= C_{B\alpha \beta \gamma \delta}^{m} + C_{K\alpha \beta \gamma \delta}^{m} + 
C_{C\alpha \beta \gamma \delta}^{m} - C_{N\alpha \beta \gamma \delta}^{m}
\\&
= C_{A\alpha \beta \gamma \delta}^{m} - C_{N\alpha \beta \gamma \delta}^{m}.
\end{aligned}
\label{localmodulus1}
\end{equation}
Here $C_{B\alpha \beta \gamma \delta}^{m}$ is the Born term, $C_{K\alpha \beta \gamma \delta}^{m}$ the kinetic contribution, $C_{C\alpha \beta \gamma \delta}^{m}$ the pressure correction~\cite{barron_1965}, and $-C_{N\alpha \beta \gamma \delta}^{m}$ the non-affine term. (Note that the Born term $C_{B \alpha \beta \gamma \delta}^{m}$ is the second derivative of the energy density with respect to the Green-Lagrange strain tensor~\cite{lutsko_1989,lutsko_1988}. Therefore, if we define the modulus by using the linear strain tensor, the stress correction term $C_{C \alpha \beta \gamma \delta}^{m}$ is necessary as long as the stress tensor has finite valued components~\cite{barron_1965}.) 

The quantity $C_{A\alpha \beta \gamma \delta}^{m}=C_{B\alpha \beta \gamma \delta}^{m} + C_{K\alpha \beta \gamma \delta}^{m} + C_{C\alpha \beta \gamma \delta}^{m}$ corresponds to the response of a system which deforms affinely at all scales~\cite{Wittmer_2013}. In contrast, $-C_{N\alpha \beta \gamma \delta}^{m}$ is a negative correction which accounts for the non-affinity of the deformation at small scales. Crystalline systems exhibit small values of $-C_{N\alpha \beta \gamma \delta}^{m}$, whereas this contribution becomes comparable in magnitude to $C_{A\alpha \beta \gamma \delta}^m$ in disordered systems~\cite{tanguy_2002}. 
\begin{figure}[t]
\centering
\includegraphics[width=0.48\textwidth]{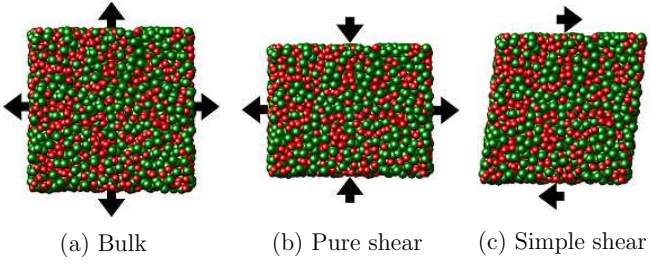}
\caption{
Schematic illustration of (a) bulk, (b) pure shear, and (c) simple shear deformations. Bulk ($K^m$), pure shear ($G_p^m$), and simple shear ($G_s^m$) local moduli corresponding to these deformations can be determined as discussed in the text.
} 
\label{eh1}
\end{figure}

The terms in Eq.~(\ref{localmodulus1}) are evaluated as:
\begin{equation}
\begin{aligned}
C_{B\alpha \beta \gamma \delta}^{m} & = \frac{1}{w^3} \Biggl< \sum_{i<j} \left( \derri{v^{ij}}{{r^{ij}}^2} - \frac{1}{r^{ij}}\deri{v^{ij}}{r^{ij}} \right) 
\\& \qquad \qquad \qquad 
\times \frac{r^{ij}_\alpha r^{ij}_\beta r^{ij}_\gamma r^{ij}_\delta}{ {r^{ij}}^2 }\frac{q^{ij}}{r^{ij}} \Biggr>, \\
C_{K\alpha \beta \gamma \delta}^{m} & = 2\left< \hat{\rho}^m \right>T (\delta_{\alpha \gamma} \delta_{\beta \delta} + \delta_{\alpha \delta}\delta_{\beta \gamma}),\\
C_{C\alpha \beta \gamma \delta}^{m} & = -\frac{1}{2} \big[ 2\left<\sigma^m_{\alpha \beta}\right>\delta_{\gamma \delta}-\left<\sigma^m_{\alpha \gamma}\right>\delta_{\beta \delta}-\left<\sigma^m_{\alpha \delta}\right>\delta_{\beta \gamma} \\
& \qquad \quad - \left<\sigma^m_{\beta \gamma}\right>\delta_{\alpha \delta} -\left<\sigma^m_{\beta \delta}\right>\delta_{\alpha \gamma} \big], \\
C_{N\alpha \beta \gamma \delta}^{m} & = \frac{V}{T} \left[ \left< \sigma^m_{\alpha \beta} \sigma_{\gamma \delta} \right>-\left< \sigma^m_{\alpha \beta} \right>\left< \sigma_{\gamma \delta} \right> \right].
\end{aligned}
\label{localmodulus2}
\end{equation}
Here, $N^m$ is the number of particles contained in the domain $m$ (dubbed $m$ hereafter), $\hat{\rho}^m=N^m/w^3$ is the local number density in $m$, $r^{ij}_\alpha$ is the vector joining particles $i$ and $j$, and $r^{ij}$ is their distance. The quantity $q^{ij}$ represents the fraction of the line segment $r^{ij}_\alpha$ which is located inside $m$. As a consequence, if $r^{ij}_\alpha$ is located outside $m$, $q^{ij}=0$, and $q^{ij}/r^{ij}$ determines the contribution of each pairwise interaction to the Born term $C_{B \alpha \beta \gamma \delta}^{m}$. Note that in principle one needs to add an impulsive correction to $C_{B \alpha \beta \gamma \delta}^{m}$ due to the truncation of the potential at the cut-off~\cite{Xu_2012}. Also we have to be careful of cut-off nonlinearities on the non-affine term $C_{N\alpha \beta \gamma \delta}^{m}$~\cite{Mizuno_2016}.
In the present case, however, we have confirmed that those correction and effect are always negligible.

Local, $\sigma_{\alpha \beta}^m$, and global, $\sigma_{\alpha \beta}$, stresses are calculated as:
\begin{equation}
\begin{aligned}
\sigma_{\alpha \beta}^m &= -\hat{\rho}^m T \delta_{\alpha \beta} + \frac{1}{w^3} \sum_{i<j} \deri{v^{ij}}{r^{ij}} \frac{r^{ij}_\alpha r^{ij}_\beta}{r^{ij}}\frac{q^{ij}}{r^{ij}}, \\
\sigma_{\alpha \beta} & = \frac{1}{V} \sum_{m} w^3 \sigma^m_{\alpha \beta}, 
\\& = -\hat{\rho} T \delta_{\alpha \beta} + \frac{1}{V} \sum_{i<j} \deri{v^{ij}}{r^{ij}} \frac{r^{ij}_\alpha r^{ij}_\beta}{r^{ij}}.
\end{aligned}
\label{localmodulus3}
\end{equation}
By using the system configurations generated by MD simulation, we can therefore directly calculate all components of ${C}^m_{\alpha \beta \gamma \delta}$ in $m$ from Eqs.~(\ref{localmodulus1})-(\ref{localmodulus3}).

{\bf Local bulk and shear moduli.} We have considered the bulk modulus, $K^m$, and the five shear moduli $G^m_l$ ($l=1,2,\cdots,5$), defined as~\cite{Mizuno_2013}:
\begin{equation}
\begin{aligned}
K^m &= (C^m_{xxxx}+C^m_{yyyy}+C^m_{zzzz}+C^m_{xxyy}+C^m_{yyxx} \\
  & \quad \ +C^m_{xxzz}+C^m_{zzxx}+C^m_{yyzz}+C^m_{zzyy})/9, \\
G^m_1 & = (C^m_{xxxx}+C^m_{yyyy}-C^m_{xxyy}-C^m_{yyxx})/4, \\
G^m_2 & = (C^m_{xxxx}+C^m_{yyyy}+4C^m_{zzzz}+C^m_{xxyy}+C^m_{yyxx} \\
    &  \quad -2C^m_{xxzz}-2C^m_{zzxx}-2C^m_{yyzz}-2C^m_{zzyy})/12, \\
G^m_3 & = C^m_{xyxy}, \\
G^m_4 & = C^m_{xzxz}, \\
G^m_5 & = C^m_{yzyz}.
\end{aligned} 
\label{equation2}
\end{equation}
The moduli $G^m_1$ and $G^m_2$ correspond to {\em pure} shear deformations (plane and tri-axial strain deformations), while $G^m_3$, $G^m_4$, and $G^m_5$ are related to {\em simple} shear deformations. We give a schematic illustration of these deformations in Fig.~\ref{eh1}. Note that the moduli defined in Eq.~(\ref{equation2}) are {\em not} eigenvalues of the modulus tensor, which is an alternative possibility~\cite{tsamados_2009,Mayr_2009}. In that case, however, the corresponding deformations, which are determined by the associated eigenvectors, are not fixed and depend on $m$. 

{\bf Distributions of the local moduli.} From the data calculated via Eq.~(\ref{equation2}), we have built the probability distribution functions, $P(C^m)$, by repetitively sampling the $20^3$ values of $C^m=K^m,\ G^m_1,\ G^m_2,\ldots,\ G^m_5$. We have confirmed that $P(C^m)$ are Gaussian distributions~\cite{yoshimoto_2004,tsamados_2009,makke_2011,Mizuno_2013} in all cases. Although $G^m_1 \neq G^m_2$ and $G^m_3 \neq G^m_4 \neq G^m_5$ in each $m$, we found $P(G^m_1)=P(G^m_2)$ and $P(G^m_3)=P(G^m_4)=P(G^m_5)$. Thus, in the following, we identify with $G^m_p$ the pure shear moduli $G^m_1$ and $G^m_2$, and with $G^m_s$ the simple shear moduli $G^m_3$, $G^m_4$, and $G^m_5$. (Note that $P(G_p^m)$ and $P(G_s^m)$ are different in cubic crystals, whereas they coincide in isotropic glasses.) 

For better clarifying a few points of our discussion, we also separately calculated from Eq.~(\ref{localmodulus1}) the affine ($C^m_A=K^m_A,\ G^m_{pA},\ G^m_{sA}$) and non-affine ($C^m_N=K^m_N,\ G^m_{pN},\ G^m_{sN}$) components of the moduli, together with the resulting $P(C^m_A)$ and $P(C^m_N)$. Finally we note that although relatively small systems ($N=4000, L=10a$) were used for these calculations, we verified that system size effects are negligible (see also Fig.~8 in Ref.~\cite{Mizuno_2013}).
\begin{figure*}[]
\centering
\includegraphics[width=0.98\textwidth]{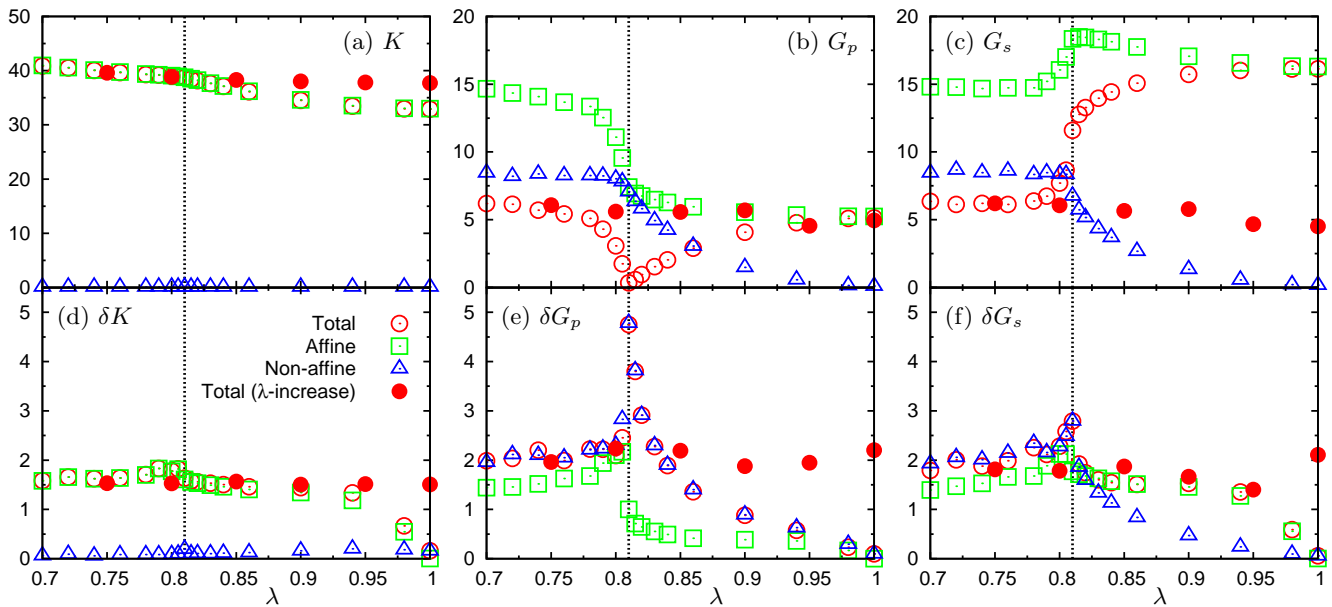}
\caption{
$\lambda$-dependence of the macroscopic (average) values of (a) bulk $K$, (b) pure shear $G_p$, and (c) simple shear $G_s$ moduli, together with the corresponding standard deviations, (d) $\delta K$, (e) $\delta G_p$, and (f) $\delta G_s$. Circles, squares, and triangles indicate the values of the total modulus $C^m$, the affine term $C^m_A$, and the non-affine term $C^m_N$, respectively. The vertical lines indicate the transition point $\lambda^\ast$, where $G_p$ vanishes and $\delta G_p \simeq 5$. We show with filled circles the data obtained by increasing $\lambda$ from $0.7$ to $1$ (only total values are shown). In this case, the local moduli distributions are insensitive to the size disorder $\lambda$ and show no significant variations in both average values and standard deviations.
} 
\label{eh}
\end{figure*}
\section{Elastic heterogeneities}
\label{sect:elastic heterogeneities}
\subsection{Disorder dependence} 
\label{sec.eh}
We have first investigated to which extent the elastic heterogeneities can be controlled by the size disorder, $\lambda$. From the distribution functions $P(C^m)$, we extracted the average values $C=K, G_p, G_s$, and standard deviations $\delta C = \delta K, \delta G_p, \delta G_s$, as:
\begin{equation}
\begin{aligned}
C &= \int C^m P(C^m) dC^m, \\
\delta C & = \sqrt{\int (C^m-C)^2 P(C^m)dC^m}.
\end{aligned}
\end{equation}
$C$ coincides with the macroscopic modulus, while $\delta C$ measures the extent of the modulus heterogeneity~\cite{Mizuno_2013}, i.e., larger values of $\delta C$ correspond to larger heterogeneities. We also calculated $C_{A(N)}$ and $\delta C_{A(N)}$ for the affine and non-affine components separately, from the distributions $P(C^m_{A(N)})$.
\begin{figure*}[]
\centering
\includegraphics[width=0.98\textwidth]{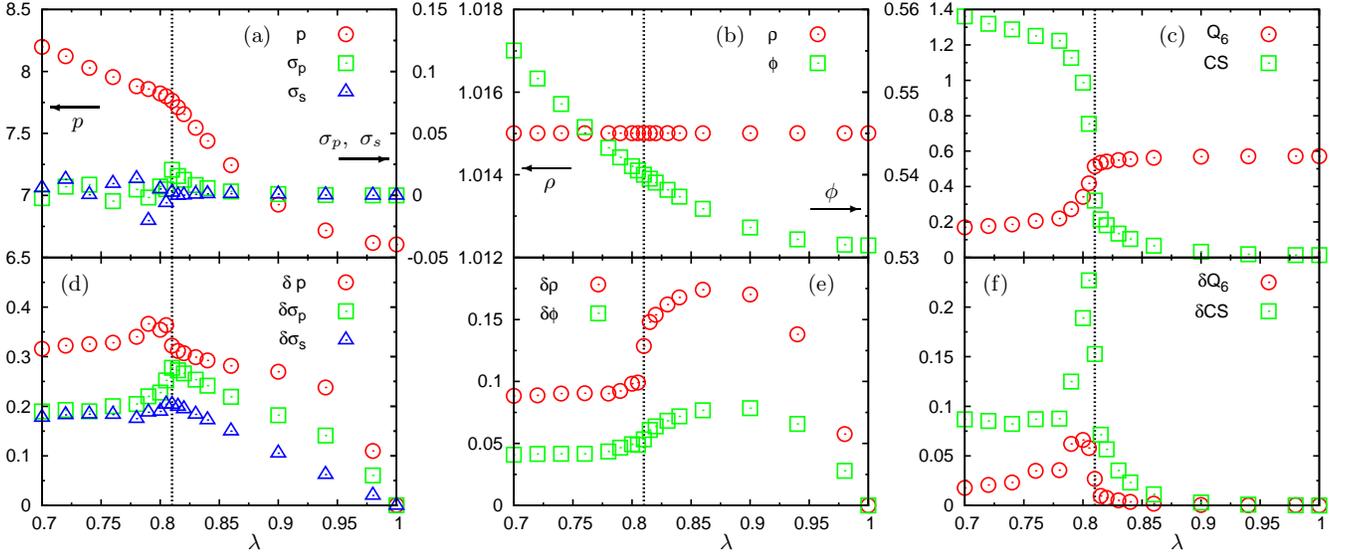}
\caption{
$\lambda$-dependence of the macroscopic (average) values of (a) pressure $p$, pure shear stress $\sigma_p$, simple shear stress $\sigma_s$; (b) mass density ${\rho}$, volume fraction $\phi$; and (c) bond-order $Q_6$ and centro-symmetry $CS$ order parameters. We also plot the corresponding standard deviations in (d) $\delta p$, $\delta \sigma_p$, $\delta \sigma_s$, (e) $\delta {\rho}$, $\delta \phi$, and (f) $\delta Q_6$, $\delta CS$. The vertical lines indicate the transition point $\lambda^\ast$. Note that the density is kept constant $\rho = 1.015$ at all values of $\lambda$, while the volume fraction mildly varies in the range of $\phi=53$ to $56\%$.
} 
\label{ehorder1}
\end{figure*}

{\bf The macroscopic moduli.} In Figs.~\ref{eh}(a)-(c), we show by open symbols the $\lambda$-dependence of $K$, $G_p$, and $G_s$, respectively, decreasing $\lambda$ (increasing the disorder) from $\lambda=1$ (perfect crystal) to $0.7$ (amorphous state). The bulk modulus $K$ assumes the highest value, the pure-shear modulus $G_p$ the lowest. In the lattice structures $G_p < G_s$, due to the affine terms $G_{pA} < G_{sA}$, whereas $G_p \simeq G_s$ in the isotropic amorphous states with $\lambda \le 0.78$. In the following we will therefore refer to $G_p^m$ and $G_s^m$ as the {\em low} and the {\em high} shear moduli, respectively.

Also, we note that the present soft-sphere model, which exhibits in the supercooled liquid state a strongly non-Arrhenius behaviour of the structural relaxation time~\cite{mizuno_2010,mizuno2_2010} and shear viscosity~\cite{mizuno_2011}, is classified as a fragile glass~\cite{Debenedetti_2001}. According to Ref.~\cite{Novikov_2004}, fragile glasses show relatively high Poisson ratios, $\nu = (3K-2G)/2(3K+G)$, compared to strong glasses. For our system, we obtain a high value, $\nu \simeq 0.43$ for $\lambda \le 0.78$, which is consistent with the findings of Ref.~\cite{Novikov_2004}. Also, fragile glasses are characterized by high atomic packing density and incompressibility~\cite{Greaves_2011}. Indeed, in the amorphous state, our system shows a high value of the bulk modulus compared to the shear modulus, $K \simeq 40 \gg G = G_p = G_s \simeq 7$.

{\bf The elastic instability.} In general, in systems with inverse-power-law interactions the non-affine component of the bulk modulus is $K_N =0$, and therefore $K = K_A$, as well demonstrated in Fig.~\ref{eh}(a). The situation is totally different for the shear moduli, $G_p$ and $G_s$, that we show in Figs.~\ref{eh}(b) and (c). For $\lambda=1$ (perfect crystal), the non-affine components are negligible, $G_{pN} \simeq G_{sN} \simeq 0$. However, as size disorder is introduced by decreasing $\lambda$, the non-affine components, $G_{pN}$ and $G_{sN}$, progressively increase. At the transition point $\lambda^\ast$, $G_{pN}$ eventually reaches the affine component $G_{pA}$, and the total $G_p$ vanishes. This observation indicates that the transition at $\lambda^\ast$ can be described as an {\em elastic instability} controlled by the low modulus $G_p$. This instability drives the structural transition, leading to steep changes of the affine terms of the shear moduli, $G_{pA}$ and $G_{sA}$, while the bulk modulus $K_A$ stays almost unchanged. For $\lambda < \lambda^\ast$, the system rapidly becomes isotropic, as manifested by the convergence $G_p \simeq G_s$. A similar instability of the shear modulus was also observed in BCC $\rightarrow$ FCC transitions of alkali metals~\cite{Milstein_1996}.

{\bf Spatial heterogeneities of local moduli.} The $\lambda$-dependence of the standard deviations, $\delta C = \delta K, \delta G_p, \delta G_s$, is shown in Figs.~\ref{eh}(d), (e), and (f) (open symbols). For $\lambda=1$ (perfect crystal), $\delta C \simeq 0$ implies $P(C^m)\approx \delta(C^m-C)$ ($\delta(x)$ is the Kronecker's delta function), i.e., the modulus is spatially homogeneous. In contrast, in the amorphous states, $\delta K \simeq 1.5$ and $\delta G_p \simeq \delta G_s \simeq 2$, implying the existence of heterogeneities in the moduli distributions.
As $\lambda$ decreases from $\lambda = 1$, the heterogeneities, $\delta K$, $\delta G_p$, and $\delta G_s$, undergo significant changes. First, as $\lambda$ decreases from $1$ to $0.9$, $\delta K$ and $\delta G_s$ increase monotonically, mainly due to the affine terms, $\delta K_A$ and $\delta G_{sA}$. On the other hand, the variation of $\delta G_p$ is less pronounced, as it is already dominated by the heterogeneity in the non-affine term, $\delta G_{pN}$. Next, as $\lambda$ approaches the transition point $\lambda^\ast$, the value of $\delta G_p$ increases dramatically, driven by the non-affine term $\delta G_{pN}$. At the transition $\lambda^\ast$, the distribution of $G^m_p$ becomes extremely heterogeneous, with a vanishing average value $G_p \simeq 0$, and a large standard deviation $\delta G_p \simeq 5$. Eventually, in the isotropic amorphous states below the transition $\lambda^\ast$, $\delta G_p$ and $\delta G_s$ rapidly converge to similar values $\delta G_{p} \simeq \delta G_{s} \simeq 2$. 

In Fig.~\ref{eh}, we also show (filled circles) our results in the case where we reversely increase $\lambda$ from $0.7$ to $1$ (see Fig.~\ref{transition1}(b)). In this case, the system keeps its initial amorphous state, and the distributions of local elastic moduli undergo no significant changes in average values, $C=K, G_p, G_s$, and standard deviations, $\delta C = \delta K, \delta G_p, \delta G_s$. This result indicates that controlling the moduli distributions by varying size disorder is rather difficult in the amorphous states. 
\begin{figure*}
\centering
\includegraphics[width=0.99\textwidth]{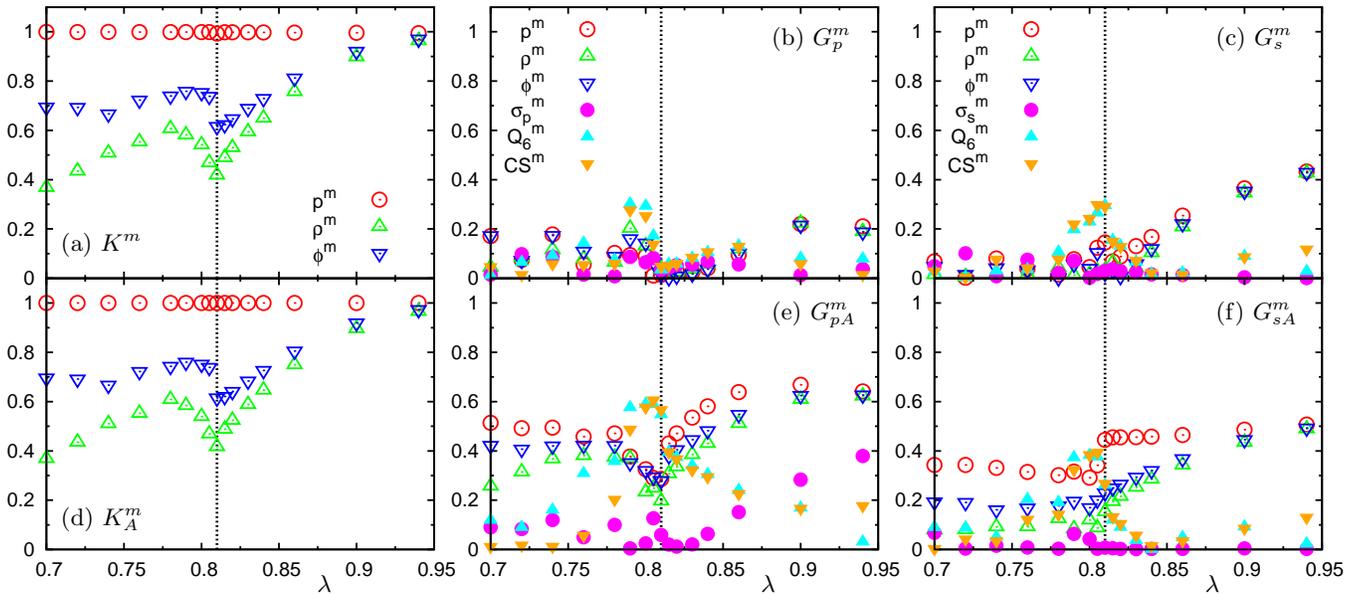}
\caption{
$\lambda$-dependence of the correlation parameters $\Psi_{C^m X^m}$ between local moduli ($C^m$ total modulus, $C^m_A$ affine term only), (a) $K^m$, (b) $G_p^m$, (c) $G_s^m$, (d) $K_A^m$, (e) $G_{pA}^m$, (f) $G_{sA}^m$, and local quantities, $X^m=p^m$, ${\rho}^m$, $\phi^m$, $\sigma_p^m$, $\sigma_s^m$, $Q_6^m$, $CS^m$. The vertical lines indicate the transition point $\lambda^\ast$. A detailed discussion of these data is included in the main text.
} 
\label{ehorder2}
\end{figure*}
\subsection{Correlation of structural quantities and elastic heterogeneities}
Disordered solids exhibit spatial heterogeneities not only in local elastic moduli, but also in other local quantities, such as local density, stress, or structural order. It is therefore interesting to try to elucidate correlations among these observables, in order to highlight the possible structural origin of elastic heterogeneities. Indeed, we may intuitively expect that values of local elastic moduli higher than the macroscopic average could be associated with denser, more close-packed regions, lower moduli to softer regions. Similarly, we could expect to observe distinct values of local elastic moduli for locally ordered structures and locally more disordered regions. 

The values of the local pressure $p^m$, and the pure $\sigma_p^m$ and simple $\sigma_s^m$ shear stresses for each $m$, were calculated from the stress tensor $\sigma^m_{\alpha \beta}$ of Eq.~(\ref{localmodulus3}), as~\cite{Mizuno_2013}:
\begin{equation}
\begin{aligned}
{p}^m &= ({\sigma^m_{xx}+\sigma^m_{yy}+\sigma^m_{zz}})/{3}, \\
{\sigma}_p^m &= ({\sigma^m_{xx}-\sigma^m_{yy}})/{2}, \quad ({\sigma^m_{xx}+\sigma^m_{yy}-2\sigma^m_{zz}})/{4}, \\
{\sigma}_s^m &= \sigma^m_{xy}, \quad \sigma^m_{xz}, \quad \sigma^m_{yz}.
\end{aligned}
\end{equation}
The mass density ${\rho}^m$, volume fraction $\phi^m$, and orientational, $Q_6^m$, and centro-symmetry, $CS^m$, order parameters were obtained from
\begin{equation}
\begin{aligned}
{\rho}^m &=\frac{1}{w^3} \sum_{i \in m} 1 = \frac{N^m}{w^3} = \hat{\rho}^m, \\
\phi^m &= \frac{1}{w^3} \sum_{i \in m} \frac{\pi}{6} (\sigma^{i})^3, \\
Q_6^m &= \frac{1}{N^m} \sum_{i \in m} q_6^i, \\
CS^m &= \frac{1}{N^m} \sum_{i \in m} cs^i.
\end{aligned}
\end{equation}
Here, $q_6^i$ and $cs^i$ are the values pertaining to particle $i$ (see for details, Refs.~\cite{Steinhardt_1983, Lechner_2008} for $q_6^i$, and Ref.~\cite{Kelchner_1998} for $cs^i$). The FCC crystal is characterized by $q_6^i \simeq 0.57$ and $cs^i \simeq 0$, whereas lower values of $q_6^i$ and higher values of $cs^i$ are expected for amorphous phases. Similarly to local moduli considered in the previous Section, we calculated the average (macroscopic) value and the standard deviation for all local quantities defined above.

{\bf The local structure.} Our results as a function of $\lambda$ are shown in Fig.~\ref{ehorder1}. At $\lambda=1$, all standard deviations assume vanishing values, i.e., the local quantities are homogeneously distributed in space. Since for inverse-power-law potentials $p^m \sim K^m$, the pressure $p^m$ shows the same heterogeneity as the bulk modulus $K^m$, i.e., $\delta p/p \simeq \delta K/K$. The heterogeneities of the shear stresses, $\delta \sigma_p$ and $\delta \sigma_s$, show a $\lambda$-dependence similar to that of $\delta p$, with an average $\sigma_p \simeq \sigma_s \simeq 0$.

In our simulations the macroscopic number density $\hat{\rho}$ and mass density $\rho = \hat{\rho}$ are kept constant, and the volume fraction $\phi$ mildly varies in the range of $\phi = 53$ to $56 \%$. Locally, however, ${\rho}^m$ and $\phi^m$ fluctuate, even in the disordered crystalline states. Indeed, in these cases, the particles are still tethered to the crystal lattice nodes, as manifested by very small values of $\left< \Delta r^2\right>$ in Fig.~\ref{transition1}(a), but they slightly deviate from the exact lattice sites positions, leading to non-zero values for $\delta {\rho}$ and $\delta \phi$.

Also, for $\lambda >0.86$ the local order parameters, $Q_6^m$ and $CS^m$, show values corresponding to those of the fcc crystalline structures, $Q_6 \simeq 0.57$ and $CS \simeq 0$, together with $\delta Q_6 \simeq 0$ and $\delta CS \simeq 0$. In contrast, as $\lambda$ approaches $\lambda^\ast$ from above, $Q_6^m$ and $CS^m$ start to fluctuate, with the respective variances strongly increasing around $\lambda^\ast$. Eventually, just below the amorphisation transition, these fluctuations keep significantly enhanced values, indicating the coexistence of lattice- and amorphous-like local environments~\cite{Hamanaka_2006,Hamanaka_2007}. In the fully developed amorphous states, $\lambda \le 0.78$, $\delta Q_6$ and $\delta CS$ converge to finite values.

{\bf Correlations.} In order to quantify the degree of correlation between the local moduli $C^m$ and the above local structural observables $X^m$, we have calculated the correlation parameters,
\begin{equation}
\Psi_{C^m X^m} = \left| \left< \left( \frac{C^m-C}{\delta C} \right) \left( \frac{X^m-X}{\delta X} \right) \right>_m \right|,
\end{equation}
where $\left< \right>_m$ is the average over all cubic domains $m$. If the variables $C^m$ and $X^m$ are {\em perfectly} correlated, we expect $\Psi_{C^m X^m} = 1$, while $\Psi_{C^m X^m} = 0$ for the perfectly uncorrelated case.~\footnote{If $C^m$ and $X^m$ are perfectly correlated, the probability distribution function of $C^m$ and $X^m$ can be written as $P(C^m,X^m)= P(C^m)\delta(C^m \pm X^m)$, implying
\begin{equation*}
\begin{aligned}
& \Psi_{C^m X^m} \\
& = \left| \int \left( \frac{C^m-C}{\delta C} \right) \left( \frac{X^m-X}{\delta X} \right) P(C^m,X^m) dC^m dX^m \right|, \\
& = \int \left( \frac{C^m-C}{\delta C} \right)^2 P(C^m) dC^m=1.
\end{aligned}
\end{equation*}
In contrast, if $C^m$ and $X^m$ have a vanishing correlation, $P(C^m,X^m) = P(C^m)P(X^m)$, and therefore $\Psi_{C^m X^m}=0$.
}
The $\lambda$-dependence of the $\Psi_{C^m X^m}$ is shown in Fig~\ref{ehorder2}, for the total moduli, $C^m =K^m, G_p^m, G_s^m$ ((a), (b), (c)), and the affine contributions alone, $C^m_A =K^m_A, G_{pA}^m, G_{sA}^m$ ((d), (e), (f)).

Since $K^m \simeq K^m_A \sim p^m$ in this case, trivially $\Psi_{K^m p^m} \simeq \Psi_{K^m_A p^m} \simeq 1$, as shown in Figs.~\ref{ehorder2}(a) and (d). $\phi^m$ can also be considered a good predictor for the bulk modulus at all $\lambda$'s, whereas the correlation with the density $\rho^m$ tends to decrease in the amorphous states, $\lambda \le 0.78$. In contrast, the shear moduli, $G_{p}^m$ and $G_{s}^m$, only show small correlations with local quantities, as shown in Figs.~\ref{ehorder2}(b) and (c). The affine terms, $G_{pA}^m$ and $G_{sA}^m$, are relatively correlated with $p^m$ and $\phi^m$ (Figs.~\ref{ehorder2}(e), (f)), as it is the bulk modulus $K^m_A \simeq K^m$. Those correlations are therefore lost due to the effect of the non-affine terms, $G_{pN}^m$ and $G_{sN}^m$. 

Two additional observations are in order. First, correlations with the order parameters $Q_6^m$ and $CS^m$, are enhanced around $\lambda^\ast$. This effect can be explained by recalling that ordered and amorphous structures, which show respectively lower (higher) and higher (lower) values of shear modulus $G^m_p$ ($G^m_s$), coexist locally. In the amorphous states ($\lambda \le 0.78$), in contrast, very small correlations only are found with $Q_6^m$ and $CS^m$. Second, it is worth to note that the two affine terms, $G_{pA}^m$ and $G_{sA}^m$, should feature very similar correlations in the isotropic amorphous structures. We have found, however, that they show different values of $\Psi_{C^m_A X^m}$, even in the deeply amorphous state $\lambda=0.7$. This result seems to indicate that some anisotropies still survive, as a memory of the initial perfect crystal structure. Although the distribution of the two affine shear moduli are very similar, $P(G_{pA}^m) \simeq P(G_{sA}^m)$ for $\lambda \le 0.78$, weak anisotropies can therefore still be detected from correlations with local quantities, even in cases where the moduli distributions are indistinguishable.

{\bf Open issues.} To summarize, $p^m$ and $\phi^m$ show clear correlations with the bulk modulus $K^m$, i.e., we indeed measured higher values of bulk modulus in denser and closely packed regions. Slight deviations of the particles positions from the perfect lattice sites induce the heterogeneities of $\delta p$ and $\delta \phi$, which are the origin of the heterogeneity $\delta K$ developing for $0.9\le\lambda\le 1$. The origin of the high shear modulus heterogeneity $\delta G_s$ can also be partially associated with the fluctuations of $p^m$ and $\phi^m$ in the disordered crystalline states. This is not the case, however, for the amorphous states. Also, we have not found any clear correlation for the shear moduli, $G_{p}^m$ and $G_{s}^m$, probably due to some subtle effect caused by the important non-affine components. The local structural origin of the shear moduli heterogeneities is therefore still an open issue~\cite{yoshimoto_2004,Tong_2014}, as it is the origin of the diverging behaviour of $\delta G_p$ as $\lambda$ approaches $\lambda^\ast$.
\begin{figure}[t]
\centering
\includegraphics[width=0.48\textwidth]{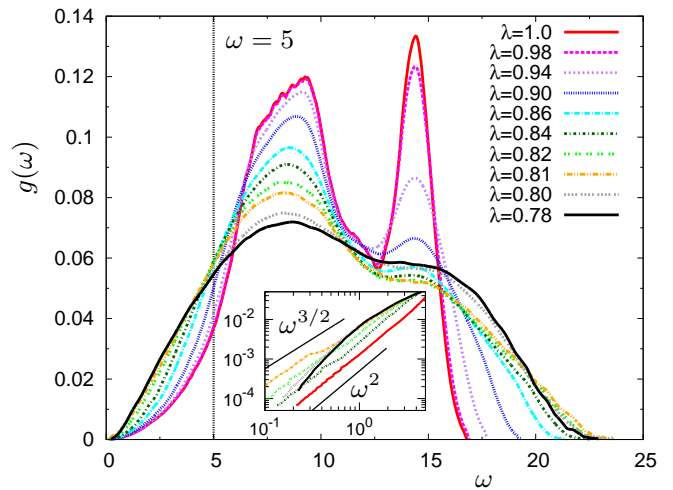}
\caption{
Vibrational densities of states, $g(\omega)$, at the indicated values of $\lambda$. These data are discussed in depth in the main text. The vertical line indicates $\omega =5$ for reference. The inset shows the details of the low-$\omega$ ($\omega < 5$) region on double-logarithmic scale. We show with lines $g(\omega) \propto \omega^2$ for the case of the perfect crystal at $\lambda =1$, and $g(\omega) \propto \omega^{3/2}$ at the amorphisation transition point $\lambda\simeq \lambda^\ast$. 
}
\label{dos}
\end{figure}
\section{Vibrational excitations} 
\label{sec.vs}
\subsection{Density of states and participation ratios}
In this Section we characterize the system vibrational states in terms of the vibrational density of states, and participation ratios and life-times of the vibrational modes. In particular, we quantify the modifications due to the modulation of the local mechanical response above and below the amorphisation transition. We also investigate the behaviour of sound-like excitations. We show that variations with $\lambda$ of the vibrational observables closely mirror the changes in the elastic heterogeneities, allowing one to establish clear correlations with different moduli for different regions of the spectrum.

\begin{figure}[t]
\centering
\includegraphics[width=0.48\textwidth]{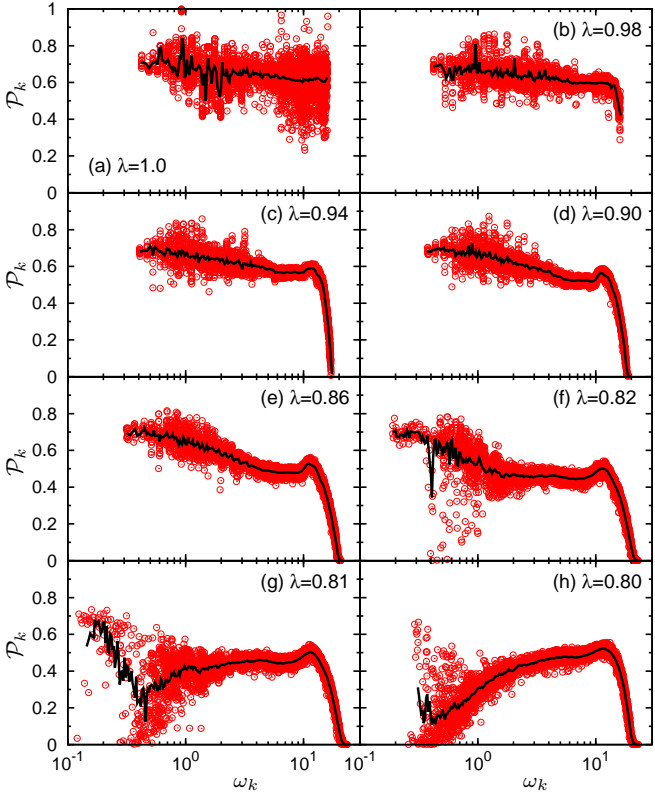}
\caption{
Participation ratios, ${\cal P}_k$, versus the eigen-frequencies, $\omega_k$, at the indicated values of $\lambda$. The solid lines represent the averaged values $\left<{\cal P}_k \right>$ calculated by smoothing the data in bins of width $\Delta \omega = 0.015$. A detailed discussion of these data is given in the main text. 
} 
\label{participation}
\end{figure}

{\bf Normal-modes analysis.} For each value of $\lambda$, we performed a standard normal modes analysis, diagonalizing the Hessian matrix calculated at the local minima of the potential energy landscape (the inherent structures)~\cite{kettel,Ashcroft,matsuoka_2012}. We have obtained the eigenvalues, $\omega_k^2$, and the corresponding eigenvectors, $\mathbf{e}^j_k$, where $j=1,\ldots,N$ and $k=1,\ldots,3(N-1)$ are the atomic and eigenmode indexes, respectively. From the histogram of the $\omega_k$ we have calculated the vDOS as
\begin{equation}
g(\omega) = \frac{1}{3(N-1)} \sum_{k=1}^{3(N-1)} \delta (\omega-\omega_k).
\end{equation}
From the eigenvectors, $\mathbf{e}^j_k$, we have calculated the participation ratios,
\begin{equation}
{\cal P}_k = \frac{1}{N} \left[ \sum_{j=1}^N ( \mathbf{e}^j_k \cdot \mathbf{e}^j_k )^2 \right]^{-1},
\label{eq:prk}
\end{equation}
which quantify the extent of localization of the vibrational mode $k$~\cite{mazzacurati_1996,Schober_2004}. As a reference, ${\cal P}_k=2/3$ for an ideal standing plane wave, and ${\cal P}_k \simeq 1/N$ for an ideal localized mode involving one particle only. For these calculations we have generated additional systems with $L$ ranging from $L=10a$ ($N=4000$) to $30a$ ($N=108000$), in order to adequately sample the lower frequency region of the spectrum~\cite{Mizuno2_2013}.
We show the $\lambda$-dependence of $g(\omega)$ in Fig.~\ref{dos}, and the data for the participation ratios ${\cal P}_k$ in Fig.~\ref{participation}. In Fig.~\ref{participation} we also plot the averaged values $\left<{\cal P}_k \right>$ (solid lines) calculated by smoothing the data in bins of width $\Delta \omega = 0.015$.

{\bf Density of states.} In the $g(\omega)$ of the perfect crystal ($\lambda=1$) we can  identify the longitudinal branch, centered around $\omega=14.5$, and the transverse branch for $\omega \in [7:9.5]$, as expected. In addition, at low frequencies $g(\omega) \propto \omega^2$ (inset of Fig.~\ref{dos}), which is consistent with the prediction of the Debye model. As $\lambda$ decreases, the above well identified phonon branches continuously loose their identity. In particular, as $\lambda$ decreases from $1$ to $0.9$, the high-$\omega$ longitudinal branch, is progressively suppressed, and a certain fraction of vibrational modes pertaining to the same branch become even localized, with low values of ${\cal P}_k$, as shown in Fig.~\ref{participation}(d) for $\lambda=0.9$. This behaviour of $g(\omega)$ and ${\cal P}_k$ have been shown to correlate with the mechanical heterogeneities associated with the bulk $\delta K$ and the high shear $\delta G_s$ moduli~\cite{Mizuno2_2013}. These quantities therefore certainly play an important role in modifying the high-$\omega$ modes, which transform from true (delocalized) phonons to more complex excitations, even localized.

\begin{figure}[b]
\centering
\includegraphics[width=0.48\textwidth]{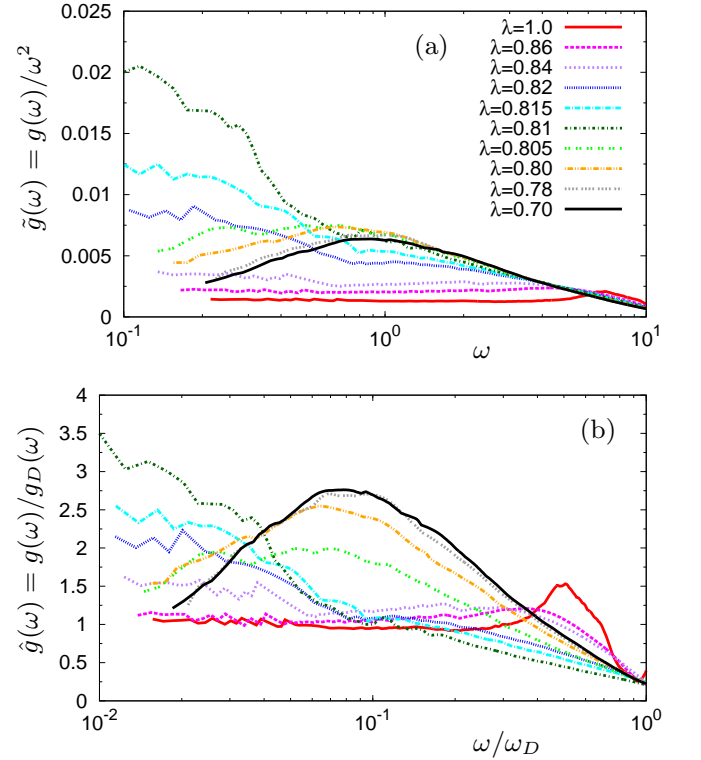}
\caption{
(a) Reduced density of state $\tilde{g}(\omega)=g(\omega)/\omega^2$ at the indicated values of $\lambda$. (b) $\hat{g}(\omega)=g(\omega)/g_D(\omega)$ plotted versus $\omega/\omega_D$, where $g_D(\omega)$ is the Debye-model prediction, and $\omega_D$ the Debye frequency. For $\lambda \le 0.78$, we observe the boson peak, at the frequency $\Omega^\text{BP} \sim 1$, typical of many studies on Lennard-Jones type of glasses~\cite{monaco2_2009,shintani_2008}.
}
\label{rdos}
\end{figure}

Next, as $\lambda$ approaches $\lambda^\ast$, where the low shear modulus $G_p^m$ fluctuates significantly around the average value $G_p \simeq 0$, the low frequency modes are increasingly populated, as indicated by the enhancement at low-$\omega$ of the reduced vDOS, $\tilde{g}(\omega)=g(\omega)/\omega^2$, shown in Fig.~\ref{rdos}(a). Interestingly, the largest value of $\tilde{g}(\omega)$ is reached at the lowest accessible frequency, possibly at $\omega \rightarrow 0$ as $\lambda \rightarrow \lambda^\ast$. Exactly at $\lambda^\ast$, we observe $g(\omega) \sim \omega^{3/2}$ (see inset of Fig.~\ref{dos}), a strongly non-Debye-like behaviour. Eventually, below the transition point $\lambda^\ast$, the reduced vDOS feature the expected BP, with $\Omega^\text{BP} \simeq 1$, already observed in glasses (see, among many others, Refs.~\cite{monaco2_2009,shintani_2008}). In addition, as $\lambda$ approaches $\lambda^\ast$, vibrational localization occurs in the low-$\omega$ region, as can be seen in Fig.~\ref{participation}(f),(g) for $\lambda=0.82$ and $0.81$ ($\lambda^\ast$). These results are clearly correlated with the behaviour of the low modulus $\delta G_p$, which therefore seems to be the relevant observable,  responsible for the modification of the low-$\omega$ part of the spectrum~\cite{Mizuno2_2013}.

\begin{figure}[t]
\centering
\includegraphics[width=0.48\textwidth]{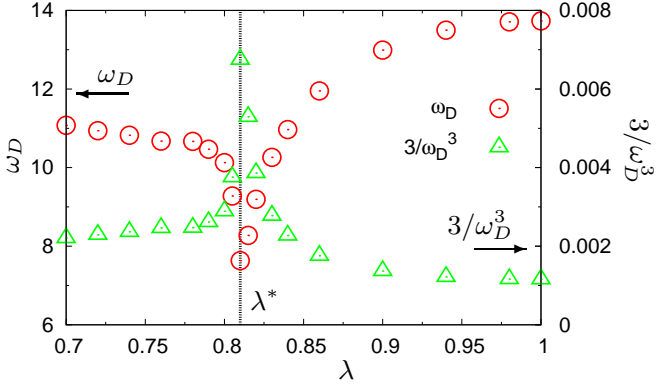}
\caption{
The Debye frequency $\omega_D$ (left axis) and the Debye level $3/\omega_D^3$ (right axis), plotted as functions of $\lambda$. The vertical line indicates the transition point $\lambda^\ast$, where $\omega_D$ and $3/\omega_D^3$ assume minimum and maximum values, respectively.}
\label{debye}
\end{figure}

{\bf A closer look at the Debye model.} To better quantify the excess of vibrational modes over the Debye model, we consider $\hat{g}(\omega)=g(\omega)/g_D(\omega)$ (Fig.~\ref{rdos}(b)), where the vDOS is scaled to the Debye-model prediction, $g_D(\omega)=\omega^2(3/\omega_D^3)$~\cite{kettel,Ashcroft,shintani_2008,monaco2_2009}. The Debye frequency $\omega_D$ and, therefore, the Debye level $3/\omega_D^3$ can be calculated directly from the macroscopic moduli $K$, $G_p$, and $G_s$. For mechanically isotropic cases with $G=G_p = G_s$, like in glasses, $\omega_D = [18 \hat{\rho} \pi^2/(1/c_L^3+2/c_T^3)]^{1/3}$, where $c_L=\sqrt{(K+4G/3)/\rho}$ and $c_T=\sqrt{G/\rho}$ are the longitudinal and transverse sound velocities, respectively. For the anisotropic case with $G_p \neq G_s$, like in cubic crystals, more complicated calculations are necessary for $\omega_D$. In this case we need to solve the Christoffel elastic equations~\cite{Jasiukiewicz_2003,Every_1979}. In Fig.~\ref{debye} we show the $\lambda$-dependence of $\omega_D$ (left axis) and $3/\omega_D^3$ (right axis). As $\lambda$ approaches $\lambda^\ast$ from above, $\omega_D$ decreases (and, consequently, $3/\omega_D^3$ is enhanced), following the decrease of the low shear modulus $G_p$. Below $\lambda^\ast$, fast converge is observed toward the values in the fully developed amorphous state, $\omega_D \simeq 11$ and $3/\omega_D^3\simeq 0.002$.

The calculated values for $\hat{g}(\omega)$ are shown in Fig.~\ref{rdos}(b) versus $\omega/\omega_D$. For $\lambda \ge 0.86$, $\hat{g}(\omega)\equiv 1$ at low frequencies, i.e., the Debye prediction holds, whereas an excess appears in $\hat{g}(\omega)$ for  $\lambda<0.86$. Previous studies have demonstrated that the $\hat{g}(\omega)$ plotted as a function of the rescaled frequency $(\omega/\Omega^\text{BP})$ collapse onto a single master curve upon increasing pressure~\cite{shintani_2008,Monaco_2006} or temperature~\cite{Caponi_2007}. In such situations, in fact, the variations of the BP can be described by a modification of the macroscopic moduli, corresponding to a global elastic transformation. In contrast, in the present case, the peak value of $\hat{g}(\omega)$ (the boson peak intensity) as well as the overall shape vary with $\lambda$, thus preventing any data collapse. This observation is similar to the results of Ref.~\cite{Niss_2007}, where the peak value of $\hat{g}(\omega)$ actually increases under increasing pressure. As already mentioned in Refs.~\cite{shintani_2008,Monaco_2006,Niss_2007}, this implies that modifications of macroscopic moduli only ({\em global} transformations) are not sufficient to fully account for the presence of the BP, and confirms that spatial distributions of local moduli ({\em local} transformations) must be considered.

{\bf More on the boson peak.} A recent study~\cite{Duval_2013} has reported that the polarization nature of the BP depends on the value of the Poisson ratio, $\nu$, the negative ratio of transverse to longitudinal strain (see above). In particular, in fragile glasses characterized by relatively high values $\nu > 0.25$, the BP has mostly a transverse origin~\cite{shintani_2008,monaco2_2009,Tan_2012}, while in strong glasses, where $\nu < 0.2$, it is of both longitudinal and transverse natures~\cite{ruffle_2006,monaco_2009,Ruffle_2010}. In this latter case, the bulk modulus features values relatively close to those of the shear modulus, and therefore both are found to affect the low-$\omega$ modes, consequently determining the nature of the BP. In our fragile system, for $\lambda<\lambda^\ast$, bulk and shear moduli are well separated ($K \simeq 40 \gg G =G_p=G_s \simeq 7$ at $\lambda \le 0.78$), and the shear modulus only can be related to the low-$\omega$ excitations in the BP region, consistently with Ref.~\cite{Duval_2013}. 

Based on the results of Fig.~\ref{rdos}(b), we can address this point more precisely. At $\lambda=0.815$ and $0.81$ ($\lambda^\ast$), where $G_p \ll G_s$ and $\delta G_p$ ($\simeq 3.5$ to $5$) is quite large, the peak values of $\hat{g}(\omega)$ are close to those at $\lambda \le 0.78$ with $G_p \simeq G_s$ and $\delta G_p \simeq \delta G_s \simeq 2$ (note that $\delta G_p + \delta G_s \simeq 4$). This observation indicates that, for $\lambda > \lambda^\ast$ where the two shear moduli are separated, only the low shear modulus heterogeneity $\delta G_p$ contributes to the excess low-$\omega$ excitations. In contrast, both (degenerate) moduli heterogeneities, $\delta G_p$ and $\delta G_s$, equally contribute below $\lambda^\ast$. We thus conclude that the lowest moduli heterogeneities are related to the BP in the entire $\lambda$-range, which can be general for disordered materials.

\subsection{Life-times of the vibrational excitations}
We now focus on the life-times of the normal modes, which are finite even in the perfect crystal phase $\lambda=1$, due to the anharmonic couplings. These finite temperature effects combine, for $\lambda<1$, with  modifications due to additional non-linearities, coming from the introduction of defects. We also clarify how these modifications impact the dynamical evolution  of the sound-like excitations propagating in the system.

{\bf The life-times of normal modes.} We can quantify the finite life-times of the normal modes as the relaxation time of the auto-correlation function $C_{E_k}(t)$ of the associated vibrational energy~\cite{McGaughey_2004,McGaughey}: 
\begin{equation}
C_{E_k}(t) = \frac{\left< \delta E_k(t) \delta E_k(0)\right>}{\left< \delta E_k^2(0) \right>},
\end{equation}
where $\delta E_k(t) = E_k(t) - \left< E_k(t) \right>$, and $E_k(t)$ is the energy of the vibrational mode $k$,
\begin{equation}
\begin{aligned}
E_k(t) &= E^\text{P}_k(t) + E^\text{K}_k(t) \\
       &= \frac{\omega_k^2}{2} S_k^\dag(t) S_k(t) + \frac{1}{2}\dot{S}_k^\dag(t) \dot{S}_k(t),
\label{modeenergy}
\end{aligned}
\end{equation}
with $\dag$ denoting complex conjugation, and
\begin{equation}
\begin{aligned}
S_k(t) &= \sum_{j=1}^N \mathbf{e}_k^j \cdot (\mathbf{r}^j(t)-\mathbf{r}_I^j), \\
\dot{S}_k(t) &= \sum_{j=1}^N \mathbf{e}_k^j \cdot \mathbf{v}^j(t).  
\label{porlization}
\end{aligned}
\end{equation}
Here, $\mathbf{e}_k^j$ is the eigenvector corresponding to the eigenfrequency $\omega_k$, $\mathbf{r}^j(t)$ and $\mathbf{v}^j(t)$ are the instantaneous position and velocity of particle $j$ at time $t$, and $\mathbf{r}_I^j$ is the position of particle $j$ in the corresponding inherent structure. For the perfect crystal ($\lambda=1$), the positions of atoms in the inherent structure coincide with the lattice sites, $\mathbf{r}_I^j \equiv \mathbf{r}^{j}_{0}$. In Eq.~(\ref{modeenergy}), $E^\text{P}_k(t)$ and $E^\text{K}_k(t)$ correspond to potential and kinetic energy of the mode $k$, respectively. Note that $\left< E_k(t) \right> = 2\left< E^\text{P}_k(t) \right>=2\left< E^\text{K}_k(t) \right>= T$, for the equipartition of energy~\cite{kettel,Ashcroft}.
\begin{figure}[t]
\centering
\includegraphics[width=0.48\textwidth]{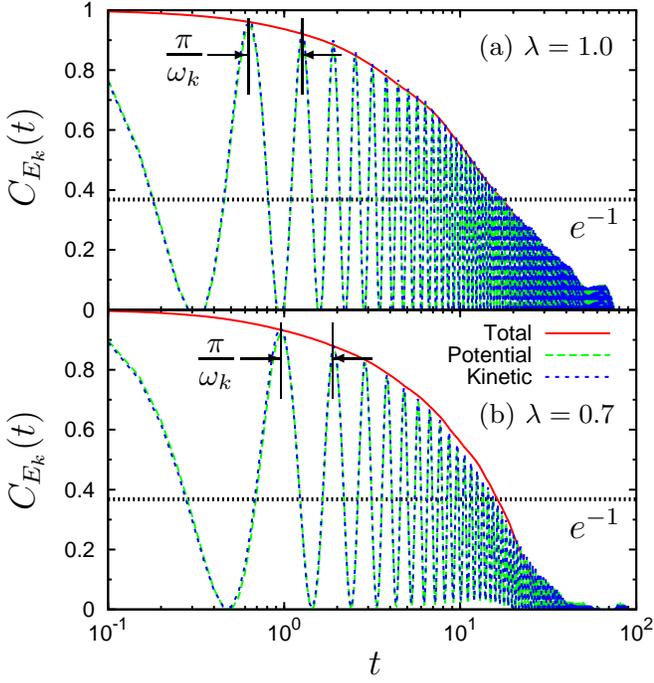}
\caption{
Auto-correlation function of the mode energy fluctuations, $C_{E_k}(t)$, for (a) $\lambda=1$, with $\omega_k=4.93$ and $\tau_k=34.6$ and (b) $\lambda=0.7$, with $\omega_k=3.27$ and $\tau_k=32.4$. Correlation functions for total $\delta E_k(t)$, potential $\delta E^\text{P}_k(t)$, and kinetic $\delta E^\text{K}_k(t)$ energies are shown. Total energy ($C_{E_k}(t)$) shows an exponential decay with a relaxation time $\tau_k/2$, where $\tau_k$ is the life-time of mode $k$. Potential ($C_{E^\text{P}_k}(t)$) and kinetic ($C_{E^\text{K}_k}(t)$) energies exhibit a dumped oscillating decay of  frequency $2\omega_k$, where $\omega_k$ is the mode frequency.
} 
\label{enecor}
\end{figure}

In Fig.~\ref{enecor} we show the temporal evolution of the energy correlation function, $C_{E_k}(t)$, for $E_k(t)$, $E^\text{P}_k(t)$, and $E^\text{K}_k(t)$, at the indicated values of $\lambda$. Both potential and kinetic energies show a damped oscillating behaviour of frequency $2 \omega_k$, whereas the total energy exhibits a simple exponential decay~\cite{McGaughey_2004,McGaughey}. The life-time $\tau_k$ of mode $k$ can be extracted as (twice) the relaxation time~\cite{McGaughey_2004,McGaughey,Ladd_1986},
\begin{equation}
\tau_k= 2 \int_0^\infty dt\; C_{E_k}(t),
\label{eq:tau1}
\end{equation}
or, equivalently,
\begin{equation}
C_{E_k}(t=\tau_k/2) = \frac{1}{e}.
\label{eq:tau2}
\end{equation}

\begin{figure}[b]
\centering
\includegraphics[width=0.48\textwidth]{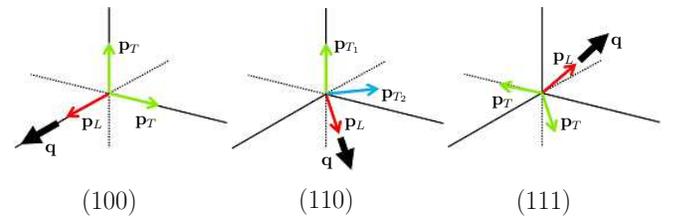}
\caption{
Schematic illustration of the considered acoustic plane waves in the three directions $(100)$, $(110)$, and $(111)$. The wave-vector $\mathbf{q}$, and the longitudinal and transverse ($X=L,T$) polarization vectors $\mathbf{p}_X$ are also shown~\cite{kettel,Ashcroft}.
}
\label{sound}
\end{figure}

{\bf Life-times of acoustic-like excitations.} Additional information comes from the life-times of acoustic-like modes~\cite{McGaughey_2004,McGaughey}, that we have studied in detail in the three propagation directions $(100)$, $(110)$, and $(111)$~\cite{Mizuno_2014}. These are defined in analogy with Eqs.~(\ref{modeenergy}) and~(\ref{porlization}) for normal modes, where we replace $\mathbf{e}_k^j$ by~\cite{McGaughey_2004,McGaughey}
\begin{equation}
\begin{aligned}
\mathbf{e}_{X \mathbf{q}}^j &= N^{-1/2} \exp(-i\mathbf{q}\cdot \mathbf{r}_I^j) \mathbf{p}_{X}.
\end{aligned}
\end{equation}
Here, the $k$-index is replaced by $X \mathbf{q}$, where $X=L,\ T\ (T_1, T_2)$, for longitudinal and transverse modes respectively, $\mathbf{q}$ is the wave-vector, and $\mathbf{p}_{X}$ is the corresponding polarization vector~\cite{kettel,Ashcroft}. The considered $\mathbf{q}$ and $\mathbf{p}_{X}$ are schematically illustrated in Fig.~\ref{sound}. The life-times can be therefore calculated from the auto-correlation function of the energy $\delta E_{X\mathbf{q}}(t)$, from relations analogous to Eqs.~(\ref{eq:tau1}) and~(\ref{eq:tau2}). Note that, since the acoustic waves are {\em not} genuine normal modes in the disordered states, the energy equipartition does not hold and $\left< E_{X\mathbf{q}}(t) \right> \neq T$.

\begin{figure*}[]
\centering
\includegraphics[width=0.995\textwidth]{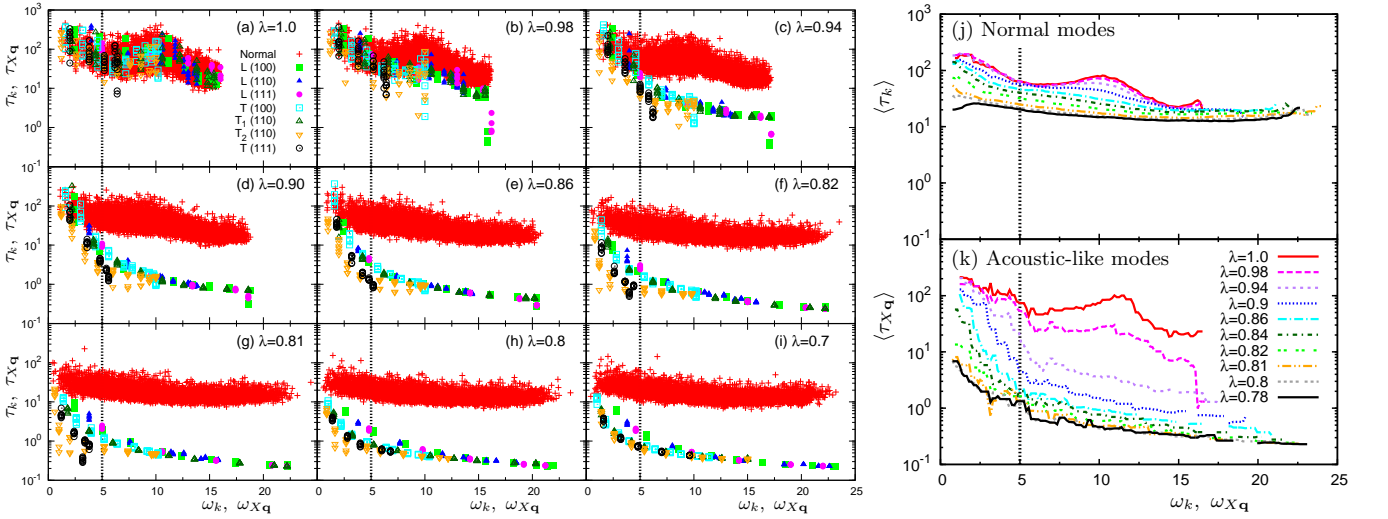}
\caption{
{\em Left panel:} Life-times of the normal modes $\tau_k$, and of the acoustic-like waves $\tau_{X\mathbf{q}}$, plotted as functions of $\omega_k$ and $\omega_{X\mathbf{q}}$, respectively, at the indicated values of $\lambda$. Here we consider the longitudinal ($L$) and transverse ($T$) acoustic waves propagating in the $(100)$, $(110)$, and $(111)$ directions (see Fig.~\ref{sound}). Specifications of the different data sets are shown in the key. The vertical lines indicate $\omega =5$ for reference. In the perfect crystalline state $\lambda=1$, $\tau_k\simeq \tau_{X\mathbf{q}}$, while in defective and  disordered states, life-times of acoustic-like modes strongly deviate from those pertaining to normal modes. For $\lambda \le 0.86$, the $\tau_{X\mathbf{q}}$ at high frequencies are of the order of the minimum time scale set by the Einstein period $\tau_E$. This is the typical time scale of the thermal motion of particles, and was estimated as $\tau_E \simeq 10^{-1}$ for the present soft-sphere system~\cite{mizuno2_2011,Mizuno_2012}. {\em Right panel:} The averaged life-times of the normal vibrational modes $\left< \tau_k \right>$ (j), and of acoustic plane waves $\left< \tau_{X\mathbf{q}} \right>$ (k), versus the frequency $\omega_k$ and $\omega_{X\mathbf{q}}$, respectively, at the indicated values of $\lambda$. These data are the same as those shown in the left panels, averaged in bins of width $\Delta \omega=1$, and over all considered propagation directions and polarizations. A comprehensive discussion of these data is included in the main text.
} 
\label{lifetimef}
\end{figure*}

{\bf Normal modes versus acoustic-like excitations.} In the left panel of Fig.~\ref{lifetimef}, we show our data sets for $\tau_k$ (normal modes) and $\tau_{X\mathbf{q}}$ (acoustic-like excitations) as a function of the corresponding frequencies, $\omega_k$ and $\omega_{X\mathbf{q}}$. For the perfect crystal at $\lambda =1$ (Fig.~\ref{lifetimef}(a)), acoustic plane waves and exact normal modes coincide, implying $\tau_{X\mathbf{q}} \simeq \tau_{k}$. Note that, as expected, data are scattered, as wave propagations are different in the three directions considered, contrary to the isotropic disordered case at $\lambda =0.7$ (Fig.~\ref{lifetimef}(i)).

As $\lambda$ decreases from $1$ to $0.7$, the life-times of the acoustic waves decrease overall of about two orders of magnitudes at any frequency, whereas normal modes show a much smaller reduction. This effect is made even more clear in the right panel of Fig.~\ref{lifetimef}, where the same data are averaged in bins of width $\Delta \omega=1$, irrespective to their longitudinal or transverse nature. Note that, for $\lambda \le 0.86$, at the higher frequencies the $\tau_{X\mathbf{q}}$ are of the order of the Einstein period, $\tau_\text{E} \sim \mathcal{O}(10^{-1})$, which is the minimum typical time-scale of thermal vibrations for the present soft-core system~\cite{mizuno2_2011,Mizuno_2012}.

\begin{figure}[t]
\centering
\includegraphics[width=0.48\textwidth]{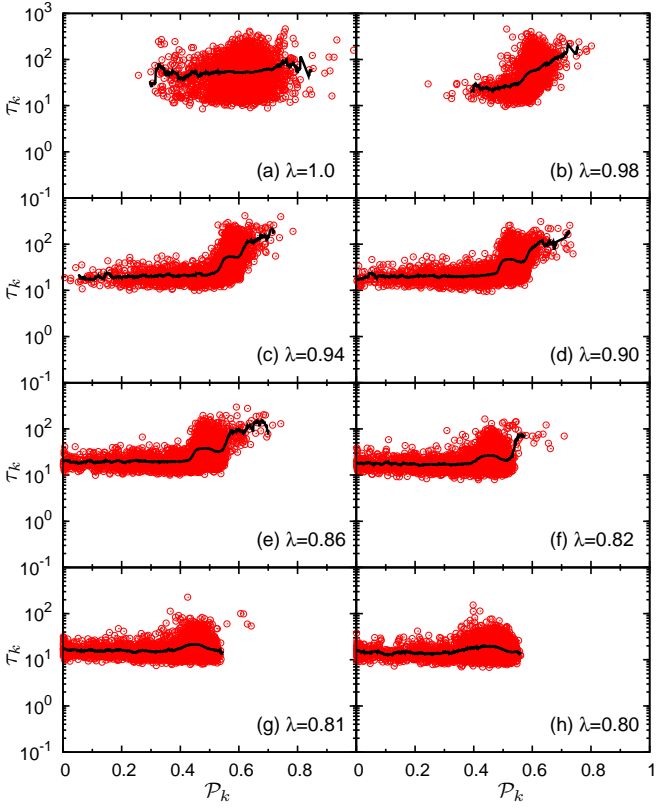}
\caption{
Parametric plot of the life-times $\tau_k$ and the participation ratios ${\cal P}_k$ for normal modes at the indicated values of $\lambda$.
The solid lines represent the averaged values $\left< \tau_k \right>$ calculated by smoothing the data in bins of width $\Delta {\cal P}_k = 0.02$.
Although modes with larger ${\cal P}_k$ indeed tend to show larger $\tau_k$, in particular in the more ordered phases, the overall correlations are quite weak. 
}
\label{lifetimepr}
\end{figure}

Additional insight on how mild variations with frequency of $\tau_k$ can induce very important modifications in $\tau_{X\mathbf{q}}$ comes from Fig.~\ref{lifetimepr}. Here we plot parametrically, for each normal mode $k$, the life-times $\tau_k$ versus the corresponding participation ratios ${\cal P}_k$. Interestingly, although at each $\lambda$ modes with larger ${\cal P}_k$ tend to show higher values of $\tau_k$ (as one could expect), the overall correlation is weak and normal modes with very similar $\tau_k$ show widely varying values of ${\cal P}_k$.  This observation seems to indicate that moduli heterogeneities impact the spatial structure of the normal modes~\cite{tanguy_2010,Derlet_2012,Mizuno2_2013} rather than simply reducing their life-times. Since acoustic plane waves are superpositions of different normal modes~\cite{taraskin_2000,matsuoka_2012}, we conclude that these modifications are the main reason for  the important frequency attenuation of the acoustic-like excitations (Fig.~\ref{lifetimef}(k)).

{\bf Life-times at high and low frequencies.} The data of Fig.~\ref{lifetimef}(k) suggest an additional observation. As $\lambda$ decreases from $1$ to $0.9$, the reduction of the life-times of the acoustic waves at fixed frequency is important at high frequencies, $\omega_{X\mathbf{q}}>5$, but relatively mild for $\omega_{X\mathbf{q}}<5$. In contrast, as $\lambda$ approaches $\lambda^\ast$, the effect is reversed and the low frequency modes ($\omega_{X\mathbf{q}}<5$) show a larger variation. These results seem again to indicate that the high and low moduli heterogeneities control high and low frequency vibrational states, respectively. Also, we emphasize that the high moduli heterogeneities, $\delta K$ and $\delta G_s$, impact a large fraction of normal modes in the broad frequency range $\omega>5$. Indeed, the integral $\int_{\omega>5} g(\omega) \, d\omega\simeq 0.95$ at $\lambda =1$, and $\simeq 0.9$ at $\lambda \le 0.8$, indicating that $90$ to $95~\%$ of the total number of normal modes is included in this frequency range. In contrast, the low modulus heterogeneity $\delta G_p$, only influences a small fraction of the spectrum with $\omega<5$, including only $5$ to $10~\%$ of the total number of normal modes. (Note that the $\Omega^\text{BP}$ is comprised in this region.)

{\bf Life-times are controlled by the heterogeneities.} In Ref.~\cite{Mizuno_2014}, we analyzed in details the attenuation rates $\Gamma_{X\mathbf{q}} \sim \tau_{X\mathbf{q}}^{-1}$ restricted to the (lowest) transverse branch of the low-frequency acoustic excitations ($\omega_{X\mathbf{q}}<5$), extracted from the line-broadening of the transverse dynamical structure factors. We also clarified their relation with the lowest shear modulus heterogeneities. Intriguingly, we found an exponential behaviour, $\Gamma_{X\mathbf{q}} \sim \exp(\delta G/g_{\tau^l})$, with $g_{\tau^l} \simeq 0.5$, $\delta G = \delta G_p$ for $\lambda \ge \lambda^\ast$, and $\delta G = \delta G_p + \delta G_s$ for $\lambda < \lambda^\ast$. We now additionally investigate this point, based on the present new data sets.

\begin{figure}[t]
\centering
\includegraphics[width=0.48\textwidth]{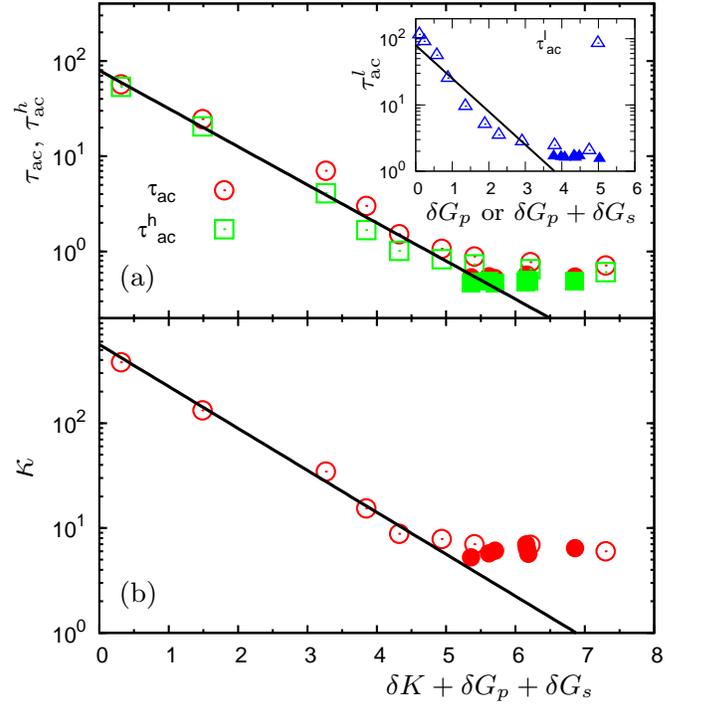}
\caption{
(a) The life-times $\tau_\text{ac}$, $\tau^h_\text{ac}$, $\tau^l_\text{ac}$ of the acoustic waves, averaged over the entire, high ($\omega>5$) and low ($\omega<5$) frequency ranges, respectively, as detailed in the text. In the main panel, the data of $\tau_\text{ac}$ and $\tau^h_\text{ac}$ are plotted as functions of the extent of the elastic heterogeneities, $\delta K +\delta G_p+\delta G_s$, for $\lambda \ge \lambda^\ast$ (open symbols) and $\lambda < \lambda^\ast$ (closed symbols). The line is an exponential fit $\tau_\text{ac} \simeq \tau^h_\text{ac} \sim \exp[-(\delta K + \delta G_p + \delta G_s)/g_{\tau}]$, with $g_{\tau}\simeq 0.4$. In the inset, we plot $\tau^l_\text{ac}$ as a function of $\delta G = \delta G_p$ for $\lambda \ge \lambda^\ast$ (open symbols) and $\delta G = \delta G_p + \delta G_s$ for $\lambda < \lambda^\ast$ (closed symbols). The solid line is a fit of the form $\tau_\text{ac}^l \sim \exp[-\delta G/g_{\tau^l}]$ with $g_{\tau^l} \simeq 0.5$. (b) The thermal conductivity $\kappa$ shown as a function of the extent of the elastic heterogeneity, together with an exponential fit $\kappa \sim \exp[-(\delta K +\delta G_p + \delta G_s)/g_\kappa]$, with $g_\kappa \simeq g_{\tau} \simeq 0.4$. Note that in the highly disordered states with large values of $\delta K + \delta G_p + \delta G_s$, both $\tau_\text{ac}$ and $\kappa$ reach the minimum allowed values, where the life-time $\tau_\text{ac} \sim \mathcal{O}(10^{-1})$ is the Einstein period~\cite{mizuno2_2011,Mizuno_2012}, and the mean-free-path of the acoustic waves is of the order of the particles diameter. 
}
\label{ehthc}
\end{figure}

Assuming that $\left< \tau_{X\mathbf{q}} \right>$ (Fig.~\ref{lifetimef}(k)) represents the typical life-time of the acoustic-like excitation of frequency $\omega$, we have determined at each $\lambda$ three (frequency-independent) life-times, $\tau_\text{ac}$, $\tau^h_\text{ac}$, and $\tau^l_\text{ac}$, averaged over the entire spectrum and, separately, in the high ($\omega>5$) and low ($\omega<5$) frequency regions. These averages obviously involve the number of sound waves comprised in those spectrum regions.
If we assume that this number does not change with $\lambda$, it is directly provided by the vDOS of the perfect crystal at $\lambda=1$ ($g_{\lambda=1}(\omega)$), where the acoustic-like excitations are the normal modes. We can, therefore, write  
\begin{equation}
\begin{aligned}
\tau_\text{ac} &= \int \left< \tau_{X\mathbf{q}} \right> g_{\lambda=1}(\omega) \,d\omega,\\
\tau^{h(l)}_\text{ac} &= \frac{ \int_{\omega>(<)5} \left< \tau_{X\mathbf{q}} \right> g_{\lambda=1}(\omega) \,d\omega }{ \int_{\omega>(<)5} g_{\lambda=1}(\omega) \,d\omega }.
\end{aligned}
\end{equation}
In Fig.~\ref{ehthc}(a) we plot $\tau_\text{ac}$ as a function of the extent of the elastic heterogeneities, $\delta K +\delta G_p+\delta G_s$, together with an exponential fit of the form (solid line),
\begin{equation}
\tau_\text{ac} \sim \exp \left(-\frac{\delta K +\delta G_p+\delta G_s}{g_{\tau}} \right).
\label{exprelation}
\end{equation}
This is similar to what we considered for the low-frequency transverse acoustic waves attenuations in Ref.~\cite{Mizuno_2014} where, however, only the shear contribution was included in the argument of the exponential. Also note that the adjusted value $g_{\tau} \simeq 0.4$ must be compared to $g_{\tau^l} \simeq 0.5$ found in Ref.~\cite{Mizuno_2014}. Since $95~\%$ of the acoustic modes are included in the high-$\omega$ region, it also results $\tau^h_\text{ac} \simeq \tau_\text{ac}$ (see Fig~\ref{ehthc}(a)). In contrast, $\tau^l_\text{ac}$ is not controlled by the total elastic heterogeneities, $\delta K +\delta G_p+\delta G_s$, but rather by the lowest one $\delta G$ only ($\delta G = \delta G_p$ for $\lambda \ge \lambda^\ast$ and $\delta G = \delta G_p + \delta G_s$ for $\lambda < \lambda^\ast$). Indeed, in the inset of Fig.~\ref{ehthc}(a) we show $\tau^l_\text{ac}$ versus $\delta G$, together with a fit of the form $\tau^l_\text{ac} \sim \exp (-\delta G/g_{\tau^l})$ with $g_{\tau^l}\simeq 0.5$, which is fully consistent with our previous observation~\cite{Mizuno_2014}.

In summary, based on the above results we propose the following scenario. By decreasing $\lambda$, as the extent of the elastic heterogeneities grows, $\tau_\text{ac}$ ($\simeq \tau^h_\text{ac}$) decreases monotonically and eventually reaches the minimum possible value, corresponding to the Einstein period, $\tau_\text{ac} \sim \mathcal{O}(10^{-1})$~\cite{mizuno2_2011,Mizuno_2012}. Although the low shear modulus $\delta G_p$ obviously exerts some influence, $\delta K$ and $\delta G_s$ turn out to be the main cause for changes  in the life-time of the acoustic excitations, influencing a predominant fraction of the vibrational modes with $\omega>5$. In contrast, low-frequency acoustic excitations with $\omega<5$, result to be only influenced by the lowest modulus heterogeneities ($\delta G_p$ or $\delta G_p + \delta G_s$ for the crystal and amorphous case, respectively)~\cite{Mizuno_2014}. The above results therefore provide us with a direct correlation between acoustic-like excitations and elastic heterogeneity~\cite{Mizuno_2014} in the entire frequency range of the vibrational excitations.
\section{Thermal conductivity} 
\label{sec.thc}
\subsection{Disorder dependence}
In this Section we explore the impact of the above studied elastic heterogeneities and vibrational excitations on the thermal conductivity, $\kappa$. $\kappa$ can be calculated by non-equilibrium simulation methods, where one applies to the system a temperature gradient or a heat current, and measures the induced heat current~\cite{Mountain_1983} or temperature gradient~\cite{Plathe_1997,Jund_1999}, respectively. These methods, however, have been demonstrated to be prone to important system size effects, especially in the case of crystals where heat carriers (phonons) have long mean-free-paths~\cite{Sellan_2010}. 

{\bf The Green-Kubo thermal conductivity.} Being aware of these limitations, in the present study we employed the equilibrium method, based on the Green-Kubo (GK) formula,
\begin{equation}
\kappa = \frac{1}{3VT^2} \int_{0}^\infty \left< \mathbf{J}(t) \cdot \mathbf{J}(0) \right> dt,
\end{equation}
where $\mathbf{J}(t)$ is the heat current vector. This formulation has been shown to provide accurate determinations of $\kappa$ in the cases of both crystals~\cite{Volz_2000,Tretiakov_2004} and amorphous solids~\cite{Lee_1991}. Also, a recent study~\cite{Kinaci_2012} reported detailed results based on the Einstein relation, which is equivalent to the GK method, while the studies of Refs.~\cite{McGaughey_2004,McGaughey,Turney2_2009} confirmed that it produces values for $\kappa$ in crystals which are consistent with those determined from the Boltzmann equation. In addition, Ref.~\cite{Tretiakov_2004} presented evidences that $\kappa$ is correctly calculated by using relatively small systems without important system size effects, even for crystals.
In the present study, we have compared the values obtained from $N=4000$ and $32000$ (larger system size), and confirmed that both values coincide well with each other.
From this observation, we concluded that $N=4000$ is large enough to exclude the system size effects on our GK calculations of $\kappa$.
(See also the discussion about system size effects for the GK method in Ref.~\cite{McGaughey}.)
\begin{figure}[t]
\centering
\includegraphics[width=0.48\textwidth]{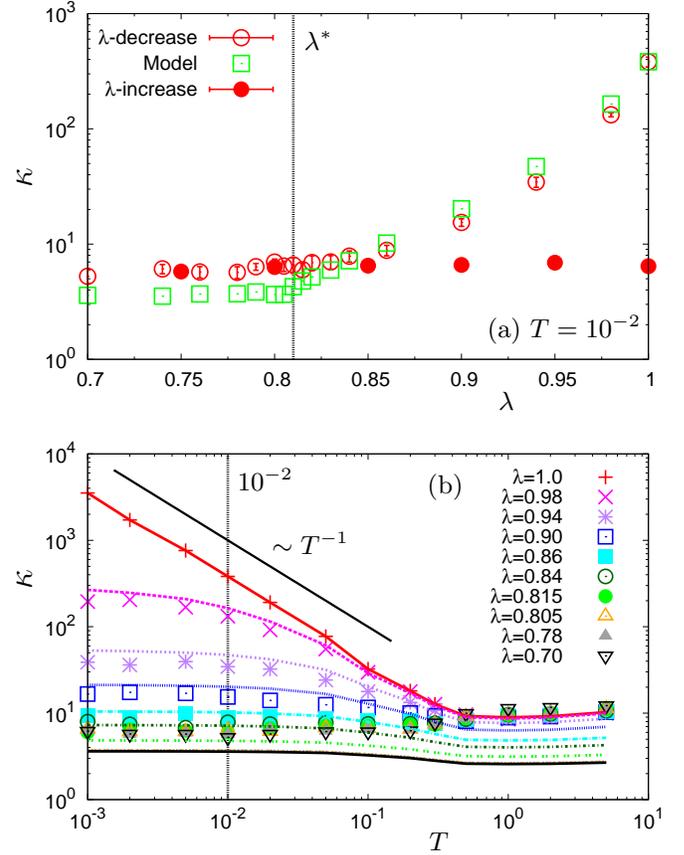}
\caption{
(a) $\lambda$-dependence of the thermal conductivity at $T=10^{-2}$ (open circles). We also show (filled circles) the data for the case where $\lambda$ is increased from $0.7$ to $1$  which show no significant variations with $\lambda$. (The system keeps the amorphous state in this case, see Fig. \ref{transition1}.) (b) $T$-dependences at the indicated values of $\lambda$. The vertical lines indicate the transition point $\lambda^\ast$ in (a), and the temperature $T=10^{-2}$ in (b). For the perfect crystal ($\lambda=1$), $\kappa$ decreases with $T$ as $\kappa \sim T^{-1}$, due to the anharmonic effects. As $\lambda$ decreases, the value of $\kappa$ decreases and saturates to a $\lambda$-independent value for $\lambda\le 0.86$. Note that $\kappa$ also becomes very mildly dependent on $T$ in the same $\lambda$-range. We also show the values of the simple model of Eq.~(\ref{kappa1}) by squares in (a) and lines in (b), which capture the overall variation of the simulation results, as detailed in the text. We recall that the melting temperature is $T_m\simeq0.6$, the glass transition temperature is $T_g \simeq 0.2$, and observe that in the liquid state, Eq.~(\ref{kappa1}) is certainly not valid. Indeed, for $T>T_m$ our simulation data for $\kappa$ converge to a value independent of $\lambda$~\cite{McGaughey}, which cannot be accounted for by the model.
}
\label{thc1}
\end{figure}

{\bf Modulating $\boldsymbol{\kappa}$ by controlling $\boldsymbol{\lambda}$.} In Fig.~\ref{thc1}(a), we show (open circles) the $\lambda$-dependence of $\kappa$, at $T=10^{-2}$. As $\lambda$ decreases from $1$ to $0.7$ (open circles), $\kappa$ is reduced by almost two orders of magnitude, similarly to the important reduction of the life-times of the acoustic-like plane waves in Fig.~\ref{lifetimef}. Indeed, the behaviour of $\kappa$ is fully consistent with that of the life-time $\tau_\text{ac}$ of the acoustic waves shown in Fig.~\ref{ehthc}(a) (compare Figs.~\ref{ehthc}(a) and (b)). The acoustic-like plane waves, rather than the normal modes, therefore play the essential role in heat conduction, for all phases.

The above evidence has two implications. First, $\kappa$ reaches the minimum allowed value already at $\lambda \simeq 0.86$, where the life-time $\tau_\text{ac}$ of the acoustic wave is of the order of the Einstein period and the mean-free-path is of the order of the particle diameter, as noted before. Second, due to $\tau_\text{ac}\simeq\tau^h_\text{ac}$ (Fig.~\ref{ehthc}(a)), we can conclude that the large number of high frequency modes ($\omega_{X\mathbf{q}}> 5$) determine $\kappa$, while those in the narrow low-frequency range ($\omega_{X\mathbf{q}} <5$) have a more limited effect. Also, in Fig.~\ref{thc1}(a) we show (closed circles) additional data for the case where $\lambda$ is reversely increased from $0.7$ to $1$ (see Fig.~\ref{transition1}). In this case, there are no significant changes by the size disorder $\lambda$ observed in $\kappa$. This is clearly correlated to the analogous behaviour of the elastic heterogeneities, as shown by the closed circles in Fig.~\ref{eh}.

{\bf The kinetic theory for $\boldsymbol{\kappa}$.} We can better characterize the thermal behaviour of the system by expressing $\kappa$ in terms of the simple kinetic theory expression~\cite{kettel,Ashcroft}:
\begin{equation}
\kappa = \frac{1}{3} \hat{\rho} C v \ell = \frac{1}{3} \hat{\rho} C v^2 \tau.
\label{kappa1}
\end{equation}
Here, $C$ is the specific heat per particle, and $v$, $\ell$, and $\tau=\ell/v$ are the average sound speed, mean-free-path, and life-time of the heat carriers (acoustic waves), respectively. Note that for our classical system the specific heat is constant, $C=3$, and the disorder (i.e., the elastic heterogeneities) influences mainly $\ell$ and $\tau$. We can reasonably assume for $\tau$ the values of $\tau_\text{ac}$ of Fig.~\ref{ehthc}(a) and, for simplicity, we also consider a constant sound speed determined as $v=(3\kappa/\hat{\rho} C \tau_\text{ac})^{1/2}\simeq 2.67$ at $\lambda=1$. We note that this value is comparable to the Debye speed of sound, $v_D =\omega_D/k_D$ ($k_D=(6\pi^2 \hat{\rho})^{1/3}$ is the Debye wave-number, $\omega_D$ are the Debye frequencies of Fig.~\ref{debye}), that assumes values in the range of $v_D \simeq 2$ to $3.5$, depending on $\lambda$. 

In Fig.~\ref{thc1}(a) we compare the simulation data with the model of Eq.~(\ref{kappa1}) (open squares), and conclude that the two data sets are in good agreement. The slight deviations for $\lambda <0.82$ very likely derive from the over-simplification of imposing a $\lambda$-independent value of $v$. The simplified models seem to capture, however, the essential features of the simulation data.

{\bf $\boldsymbol{\kappa}$ is controlled by the heterogeneity.} This agreement indirectly supports the conclusion that elastic heterogeneities significantly modify both the life-times of sound waves and the thermal conductivity. More specifically, the high modulus heterogeneities, $\delta K$ and $\delta G_s$, influence the large fraction of acoustic waves ($\omega_{X\mathbf{q}} >5$), causing the steep decrease of $\kappa$ for $\lambda>\lambda^\ast$, while for $\lambda \simeq \lambda^\ast$ the low shear modulus heterogeneity $\delta G_p$ also comes into play, affecting, however, only the narrow low-frequency regime ($\omega_{X\mathbf{q}} <5$), inducing a very small additional variation of $\kappa$. This conclusion is evident from the representation of our data shown in Fig.~\ref{ehthc}(b), where we plot $\kappa$ versus the extent of the elastic heterogeneities, $\delta K + \delta G_p + \delta G_s$. This curve seems to follow an exponential relation, $\kappa \sim \exp[-(\delta K +\delta G_p + \delta G_s)/g_\kappa]$, similar to Eq.~(\ref{exprelation}) for $\tau_\text{ac}$, with an identical value of the parameter $g_\kappa \simeq g_{\tau} \simeq 0.4$.
\begin{figure}[t]
\centering
\includegraphics[width=0.48\textwidth]{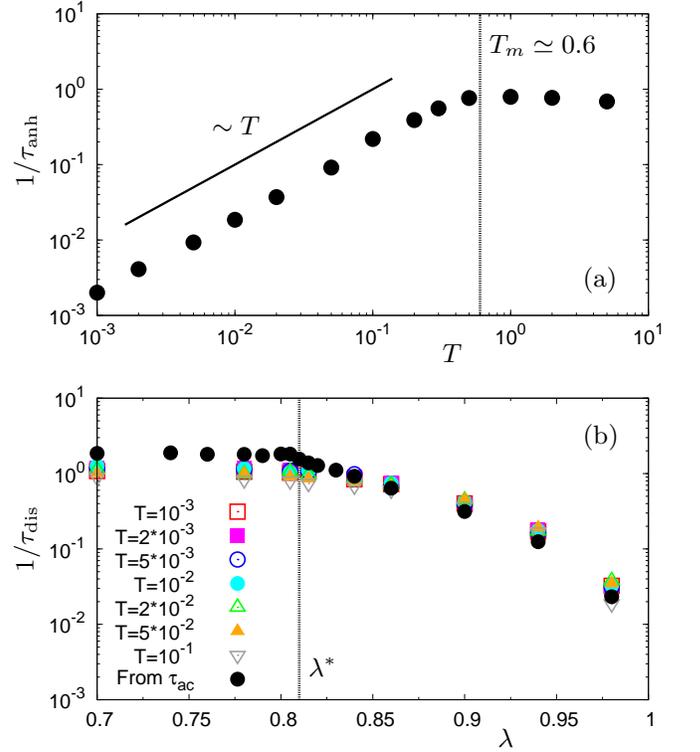}
\caption{
(a) $T$-dependence of the anharmonic term, $1/\tau_\text{anh}$ (Eq.(\ref{kappa4})) at $\lambda=1$, and (b) $\lambda$-dependence of the disorder term, $1/\tau_\text{dis}$ (Eq.(\ref{kappa3})) at the indicated values of $T$. The vertical line indicates the melting temperature $T_m \simeq 0.6$ at $\lambda=1$ in (a), and the transition point $\lambda^\ast$ in (b). In (b) we also plot (closed black circles) the values extracted from the life-times $\tau_\text{ac}$ shown in Fig.~\ref{ehthc}(a) by using Eq.(\ref{kappa5}). We can see that $1/\tau_\text{dis}$ does not depend on temperature for $T\le 10^{-1}$, and there is a good agreement with the values extracted from $\tau_\text{ac}$. This observation supports the validity of the additive decomposition of the attenuation of Eq.~(\ref{kappa2}).
} 
\label{thc1a}
\end{figure}
\subsection{Temperature dependence}
We now analyse the interplay of disorder and temperature in determining the thermal properties of the model. Our temperature data at the indicated values of $\lambda$ are shown in Fig.~\ref{thc1}(b) by symbols. We first observe that data at all values of $\lambda$ are superimposed for $T>T_m$, due  to the fact that in the liquid state size heterogeneity plays very little role in transport properties. Next, in the crystal reference state $\lambda=1$, we expect a vanishing effect ascribed to disorder and a non-trivial behaviour entirely associated with the effect of anharmonicities. This is indeed well demonstrated by the data, where at low temperatures $\kappa \sim T^{-1}$, as expected. As $\lambda$ decreases, in contrast, the effect of the locally heterogeneous elastic response becomes increasingly important, and generates a more complex reduction of $\kappa$ at all temperatures, which does not follow the simple anharmonic prediction. Eventually, $\kappa$ undergoes very mild variations with $T$ for $\lambda \le 0.86$, indicating that disorder dominates over anharmonic couplings, reducing $\kappa$ to a $T$ and $\lambda$ independent minimum value in the amorphous states.

{\bf Separating disorder and anharmonicities.} The above peculiar $T$ and $\lambda$ dependences can be described in terms of an obvious generalization of Eq.~(\ref{kappa1}), with constant $C=3$ and $v=2.67$, but $\tau=\tau(T,\lambda)$. We now assume that the attenuation rate (the inverse of the life-time) can be decomposed into two terms, as:
\begin{equation}
\frac{1}{\tau(T,\lambda)} = \frac{1}{\tau_\text{anh}(T)} + \frac{1}{\tau_\text{dis}(\lambda)}. \label{kappa2}
\end{equation}
Here, $1/\tau_\text{anh}(T)$ encodes the purely anharmonic attenuation, whereas $1/\tau_\text{dis}(\lambda)$ describes that originating from the presence of the elastic heterogeneities. We can evaluate the anharmonic attenuation by using the value of $\kappa$ for the pure crystal ($\lambda=1$),
\begin{equation}
\frac{1}{\tau_\text{anh}(T)}=\frac{\hat{\rho} C v^2}{3\kappa(T,\lambda=1)} \sim T. \label{kappa4}
\end{equation}
The corresponding data are shown in Fig.~\ref{thc1a}(a), with ${1}/\tau_\text{anh}(T)$ increasing with $T$ at low-$T$ and saturating to a constant value around the melting temperature $T_m\simeq 0.6$. As a consequence, we can extract the disorder-related term from
\begin{equation}
\begin{aligned}
\frac{1}{\tau_\text{dis}(\lambda)} &= \frac{\hat{\rho} C v^2}{3\kappa(T,\lambda)} - \frac{1}{\tau_\text{anh}(T)} \\
                                   &= \frac{\hat{\rho} C v^2}{3}  \left[\frac{1}{\kappa(T,\lambda)} - \frac{1}{\kappa(T,\lambda=1)}\right].  \label{kappa3}
\end{aligned}
\end{equation}

{\bf The effect of disorder.} If the additive decomposition of Eq.~(\ref{kappa2}) is valid, the right-hand side of Eq.~(\ref{kappa3}) should be independent of $T$. This is confirmed by the data of Fig.~\ref{thc1a}(b), where we plot $1/\tau_\text{dis}(\lambda)$ at the indicated values of $T$. These data superimpose at all temperatures, thus corroborating our hypothesis. In the same figure, we also plot (closed circles) an alternative determination of $1/\tau_\text{dis}$, based on the calculated values of $\tau_\text{ac}$ (Fig.~\ref{ehthc}(a)):
\begin{equation}
\begin{aligned}
\frac{1}{\tau_\text{dis}(\lambda)} &= \frac{1}{\tau_\text{ac}(T=10^{-2},\lambda)}-\frac{1}{\tau_\text{ac}(T=10^{-2},\lambda=1)}, \label{kappa5}
\end{aligned}
\end{equation}
where the right hand side is calculated for $T=10^{-2}$.
The two sets of data (Eqs.~(\ref{kappa3}) and~(\ref{kappa5})) are in very good agreement for $\lambda>\lambda^\ast$, while they show some discrepancies below $\lambda^\ast$. This can, again, be ascribed to the over-simplified hypothesis of a constant value for $v$. The overall similarities are however striking, confirming the strict correlation existing between the behaviour of the acoustic-like modes and heat transport.

{\bf Modelling simulation data.} In Fig.~\ref{thc1}(b), we now compare the simulation data with a model (lines) based on Eqs.~(\ref{kappa1}) and (\ref{kappa2}), where the values of $\tau_\text{anh}(T)$ and $\tau_\text{dis}(\lambda)$ are given by Eqs.~(\ref{kappa4}) and (\ref{kappa5}), respectively. The model is overall capable to capture the main features of both the $T$- and $\lambda$-dependences of $\kappa$. We have shown that $1/\tau_\text{anh}(T)$ increases with $T$, while $1/\tau_\text{dis}(\lambda)$ is enhanced as $\lambda$ decreases. The competition between these two terms finally controls the overall behaviour of $\kappa$. Here we recall the melting temperature, $T_m \simeq0.6$, and the glass transition temperature, $T_g \simeq 0.2$. In the liquid state, Eq.~(\ref{kappa1}) is certainly not valid. Indeed, for $T>T_m$ our simulation data for $\kappa$ converge to a value independent of $\lambda$~\cite{McGaughey}, which cannot be accounted for by the model. 
\begin{figure}[t]
\centering
\includegraphics[width=0.48\textwidth]{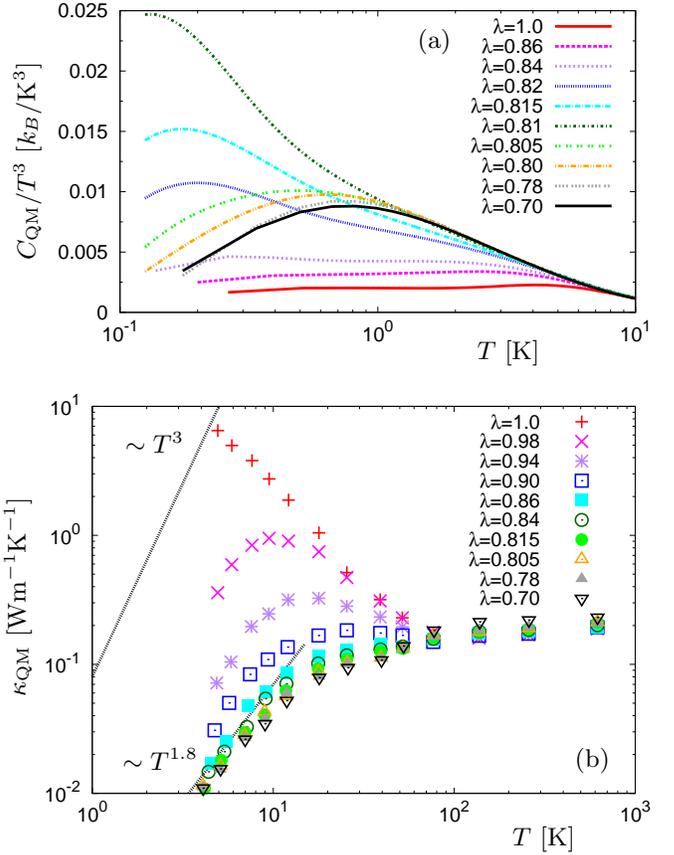}
\caption{
Temperature dependence of (a) $C_\text{QM}/T^3$ and (b) $\kappa_\text{QM}$ at the indicated values of $\lambda$.
The specific heat $C_{\text{QM}}$ (Eq.~(\ref{spheat})) is calculated from the $g(\omega)$ data of Fig. \ref{dos}.
The thermal conductivity $\kappa_\text{QM}$ is obtained by applying Eq.~(\ref{eqkappareal}) to the MD data of Fig. \ref{thc1}(b).
For those calculations, we have used physical Argon units, $\sigma =3.405~\mathrm{\AA}$, $\epsilon/k_B = 125.2$~K, and $\tau = 2.11$~ps.
In the figures, the temperature, specific heat, and thermal conductivity are measured in units of K, $k_B$, and $\text{W}\text{m}^{-1}\text{K}^{-1}$, respectively.
For $\lambda \le 0.86$, $\kappa_\text{QM}$ increases as $\kappa_\text{real} \propto T^{\gamma}$ with $\gamma=1.8\simeq 2$, and a plateau value is observed around $T \sim 20$~K. The excess peak in $C_{\text{QM}}/T^3$, however, is found in the lower temperature region $T \le 1$~K. A discussion of this point is included in the text.
} 
\label{thc2}
\end{figure}
\subsection{Quantum corrections}
\label{QC}
We now discuss the modifications to our calculations of thermal conductivity when we consider effective quantum corrections. Indeed, in classical systems including those studied here, vibrational modes of any energy are populated with the same statistical weight, conforming to a flat probability distribution. This is, however, in contrast with the principles of quantum mechanics, where a vibrational mode of frequency $\omega$ is excited according to the Bose-Einstein distribution~\cite{kettel,Ashcroft},
\begin{equation}
\begin{aligned}
f(\omega,T) &= \frac{1}{\exp(\beta\hbar\omega)-1}.
\end{aligned} \label{bedis}
\end{equation}
Here, $\beta=1/k_B T$, and $\hbar = h/2\pi$ with $h$ the Plank constant. The peculiar form of Eq.~(\ref{bedis}) has the important implication that the lower frequency modes are more excited than those pertaining to the upper part of the spectrum, modifying both the specific heat and thermal conductivity in the low-$T$ regime. As a consequence, the (per particle) specific heat $C$, which based on the equipartition theorem is a $T$-independent constant $C=3\ [k_B]$ for classical systems, depends on $T$ in the quantum formulation.

{\bf Quantum-like specific heat from classical data.} By using  the vibrational density of states of Fig.~\ref{dos}, we can approximately embed the effect of quantum correction in our calculations, and determine the quantum-like value of $C(T)$ as~\cite{kettel,Ashcroft}:
\begin{equation}
\begin{aligned}
{C_\text{QM}(T)} &= 3 \int \hbar \omega \frac{\partial f(\omega,T)}{\partial T} g(\omega) d\omega \\
&= 3 k_B \int \frac{ (\beta\hbar \omega)^2 \exp(\beta\hbar \omega)}{(\exp(\beta\hbar \omega)-1)^2} g(\omega) d\omega.
\end{aligned} \label{spheat}
\end{equation}
The Debye model $g_D(\omega) = (3/\omega_D^3)\omega^2 \sim \omega^2$ implies a specific heat $C_D(T) = (12\pi^4 k_B/5)(T/T_D)^3 \sim T^3$, with $T_D = \hbar \omega_D/k_B$ the Debye temperature. We can underline the variation of $C_\text{QM}(T)$ compared to the Debye prediction by plotting $C_\text{QM}(T)/T^3$ against $T$ in Fig.~\ref{thc2}(a). Here we have used physical Argon units, $\sigma =3.405~\mathrm{\AA}$, $\epsilon/k_B = 125.2$~K, and $\tau = 2.11$~ps. From these data we clearly see that the excess values in $g(\omega)/\omega^2$ (Fig.~\ref{rdos}(a)) are directly mirrored on the non-monotonic $T$-dependence of $C_\text{QM}(T)/T^3$, one of the main features of glasses~\cite{lowtem,Schober_1996,Zeller_1971}. In the crystal state, $\lambda=1$, the Debye prediction holds and $C_\text{QM}(T)/T^3$ is therefore $T$-independent. In contrast, as $\lambda$ tends to $\lambda^\ast$, we observe the appearance of clear maxima of increasing intensity at decreasing values of $T$. For $\lambda<\lambda^\ast$, an opposite behaviour is observed, with a rapid convergence to the final stable value in the amorphous states.

{\bf Quantum-like $\boldsymbol{\kappa}$.} This non-trivial temperature dependence of the specific heat $C_\text{QM}(T)$ must be followed by significant modifications in the $T$-dependence of the thermal conductivity. By keeping in Eq.~(\ref{kappa1}) the classical values for $v$ and $\tau$, but replacing $C$ by $C_\text{QM}(T)$, we can map the classical values $\kappa(T_\text{MD})$ to the quantum-like values $\kappa_\text{QM}(T)$ as~\cite{Lee_1991,Volz_2000,McGaughey}
\begin{equation}
\begin{aligned}
{\kappa_\text{QM}(T)} &= \left [\frac{C_\text{QM}(T)}{C}\right]\kappa(T_\text{MD}).
\label{eqkappareal}
\end{aligned}
\end{equation}
Here, $T_\text{MD}$ is the classical heat bath temperature, determined from the particles kinetic energy in the MD simulation, which we can map onto an appropriate quantum-like value, by equating the total vibrational energy of the classical and quantum systems, as
\begin{equation}
\begin{aligned}
k_B T_{\text{MD}} & = \int \hbar \omega \left[ \frac{1}{2} + f(\omega,T) \right] g(\omega) d\omega.
\end{aligned} \label{qc1}
\end{equation}
Here, the first term of the right-hand side is the zero-point energy which we excluded from our calculations, following previous works~\cite{Mizuno2_2013,Lee_1991,Volz_2000,McGaughey}. (See Ref.~\cite{McGaughey} for further details on this point.) 

{\bf Reproducing the experimental $\boldsymbol{\kappa(T)}$.} We show the temperature dependence of $\kappa_\text{QM}(T)$ in Fig.~\ref{thc2}(b). For $\lambda=1$, the computed values are similar to those determined experimentally for solid Argon~\cite{Christen_1975}, confirming the validity of our approach. Note, however, that the $T^{-1}$ dependence shown by our data does not cross-over to the predicted $T^3$ behaviour at very low temperatures. This regime is indeed expected in perfect crystals~\cite{Zeller_1971,Cahill_1988,Cahill_1992,Christen_1975}, where the mean-free-path of heat carriers cannot grow indefinitely and must eventually be limited by the finite-size of the material sample~\cite{Thacher_1967}. This discrepancy can be rationalized by noticing that, using periodic boundary conditions, we consider a system which is virtually of infinite extent in all directions. In these conditions, as $T$ decreases the mean-free-path can increase indefinitely, determining the observed non-bounded behaviour of $\kappa_\text{QM}(T)$. 

Interestingly, however, introducing a very limited amount of disorder ($\lambda=0.98$) is sufficient to trigger an inversion of the monotonicity at $T\simeq 10$~K, with the appearance of a well-defined maximum. Eventually, for $\lambda \le 0.86$, we approximately recover the $T$-dependence typical of glasses, $\kappa \sim T^\gamma$, with $\gamma=1.8 \simeq 2$, followed by a plateau value appearing around $T\simeq 20$~K. It is worth to emphasize that this glass-like behaviour is already acquired above the amorphisation transition, in the defective crystalline states. This observation is consistent with the experimental work of Refs.~\cite{Cahill_1988,Cahill_1992}, which reported a glass-like $T$-dependence of thermal conductivity for disordered crystals of mixed alkali halides and cyanides [$(\text{KBr})_{1-x}(\text{KCN})_x$, $(\text{NaCl})_{1-x}(\text{NaCN})_x$] and fluorite structure crystals [$\text{Zr}_{1-x}\text{Y}_x\text{O}_{2-x/2}$, $\text{Ba}_{1-x}\text{La}_x\text{F}_{2+x}$], concluding that disorder can produce a glass-like thermal conductivity even in positionally ordered crystals. We remark, however, that in the former case, beside size or mass disorder, librations of CN molecules are also expected to strongly couple to acoustic excitations, contributing to strong scattering and reduction of thermal conductivity~\cite{Grannan_1988}. Similarly, vacancies or interstitials can play a role similar to that as disorder in fluorite structure crystals. Different mechanisms can, therefore, contribute to achieve glass-like $\kappa$ similar as that observed here. 

{\bf The Boson peak and $\boldsymbol{\kappa(T)}$.} An observation is in order at this point. Although the values of $\kappa_\text{QM}(T)$ are consistent with earlier experimental results~\cite{Zeller_1971,Cahill_1988,Cahill_1992,Christen_1975}, one should note that the above method to include effective quantum corrections does {\em not} allow to precisely implement the Bose-Einstein distribution, as discussed in Ref.~\cite{Turney_2009}. The considered quantum correction is in fact {\em global}, in the sense that it is based on the simple expression of Eq.~(\ref{kappa1}), where a single effective excitation represents the average effect of all heat carriers. In order to properly deal with the Bose-Einstein distribution, it is necessary to consider the mode-by-mode expression,
\begin{equation}
\kappa = \frac{1}{3V} \sum_k C_k v^2_k \tau_k = \frac{1}{3V} \sum_k \hbar \omega_k \frac{\partial f(\omega_k,T)}{\partial T} v^2_k \tau_k, \label{wavebywave}
\end{equation}
where $C_k$, $\omega_k$, $v_k$, and $\tau_k$ refer to the single mode $k$~\cite{Turney_2009}. In this expression the Bose-Einstein distribution is explicitly included into the specific heat $C_k$, which makes the contribution of lower frequency modes to $\kappa$ higher. This point is not taken into account in our calculations, with the following important consequence. 

In Fig.~\ref{thc2}(b) we show that $\kappa_\text{QM}$ becomes $\lambda$-independent for $\lambda \le 0.86$, similarly to the classical $\kappa$ shown in Fig.~\ref{thc1}(b). This implies that both formulations are controlled by the high frequency modes ($\omega>5$) only, without any important contributions arising from the low-frequency excitations ($\omega <5$), including those pertaining to the BP. Indeed, Figs.~\ref{thc2}(a) and (b) indicate that the BP temperature, where $C_\text{QM}/T^3$ has a maximum, and the temperature where the plateau manifests in $\kappa_\text{QM}$, do not coincide. One should therefore conclude that these two features are {\em not} related one to the other. This conclusion, however, partially originates from the classical nature of the system, and might be modified by a correct calculation of $\kappa_\text{QM}$ based on Eq.~(\ref{wavebywave}). In this case, one should observe non-negligible overall variation of $\kappa_\text{QM}$ in the amorphous states close to the transition $\lambda^\ast$, influenced by the low-frequency vibrational excitations. 

{\bf More on quantum corrections.}  One crucial problem in applying the mode-based correction of Eq.~(\ref{wavebywave}) to disorder solids is, however, that the heat carriers (acoustic-like modes) are inconsistent with the {\em actual} normal modes of vibration. To overcome this difficulty, Allen and Feldman (AF)~\cite{Allen_1993,Feldman_1993} have proposed the alternative formulation
\begin{equation}
\kappa(T) = \frac{1}{V} \sum_k C_k D_k = \frac{1}{V} \sum_k \hbar \omega_k \frac{\partial f(\omega_k,T)}{\partial T} D_k.
\end{equation}
Here the actual normal modes $k$ carry heat with a diffusivity $D_k$, formulated on the basis of a GK formalism. This method has been successfully applied to the calculation of $\kappa$ in jammed solids~\cite{Xu_2009,Vitelli_2010}. More recently, motivated by the AF work, a method has been developed in Ref.~\cite{Wei_2016} for a direct calculation of the modal contributions to thermal conductivity, by combining the GK formula and normal mode analysis. Finally, an alternative possibility has been proposed recently, based on quasi-quantum MD simulations employing quantum thermal baths~\cite{Dammak_2009,Barrat_2011,Bedoya_2014}. In Refs.~\cite{Dammak_2009,Barrat_2011}, a good reproduction of the temperature dependence of the specific heat $C_\text{QM}(T)$ has been demonstrated, without any corrections. Unfortunately, a proper calculation of thermal conductivity is still problematic and poses severe issues~\cite{Bedoya_2014}.
\section{Conclusions and remarks}
\label{sect:conclusions}
We have investigated the interplay among local heterogeneous mechanical response, vibrational excitations, and heat transport, for a numerical model able to interpolate continuously from the perfect crystal, through increasingly defective crystalline systems, to plainly amorphous phases. By substantially improving the data sets investigated in our previous works~\cite{Mizuno2_2013,Mizuno_2014}, we have provided a general discussion in a unique framework, unifying in a single picture large part of the possible solid states of matter. In particular, by generating extremely extended ensembles of system configurations, we have: {\em i)} determined the extent of the elastic constants heterogeneities (bulk and shear moduli), and investigated possible correlations with more immediate structural features; {\em ii)} characterized in details the elementary vibrational excitations in terms of eigenvalues and eigenvectors of the Hessian matrix together with the associated life-times; {\em iii)} investigated the more involved acoustic-like excitations, as those detected in inelastic X-rays scattering experiments, for instance; {\em iv)} determined temperature and disorder dependence of thermal conductivity, with an in-depth discussion of the limitations imposed by plainly classical calculations. Take-home messages of our work include:

{\em 1.} Spatial fluctuations in local elastic moduli modify the overall structure of the vibrational modes, transforming plane waves into more complex vibrational excitations, rather than simply reducing their life-times. A substantial fraction of normal modes is also transformed in localized excitations. The above important modifications lead to a large reduction of the life-times of the acoustic-like excitations, which are superpositions of several different normal modes with different frequencies.

{\em 2.} The heterogeneity of the higher-valued moduli impact the high frequency vibrational modes, whereas the low-$\omega$ excitations are primarily modified by the heterogeneity associated to the lower-valued moduli. More precisely, the low-$\omega$ vibrational excess, identifying the boson peak in the glassy phases, is determined by the pure shear modulus $\delta G_p$ for $\lambda>\lambda^\ast$, and by the two degenerate shear moduli, $\delta G_p \simeq\delta G_s$, in the amorphous states.

{\em 3.} The acoustic plane waves play an essential role in heat conduction even in disordered solids, the thermal conductivity being related to their life-time, $\kappa \sim \tau$. The temperature, $T$, and disorder, $\lambda$, dependences of $\kappa$ are well described by a simple model based on Eqs.~(\ref{kappa1}) and~(\ref{kappa2}). This successfully reproduces the interplay between anharmonic couplings and the effect of disorder due to the presence of the elastic heterogeneities.

{\em 4.} The thermal conductivity $\kappa$ is determined by the high-$\omega$ modes ($\omega>5$), which cover most part ($90$ to $95~\%$) of the vibrational spectrum and are mainly controlled by the high moduli heterogeneities, $\delta K$ and $\delta G_s$. $\kappa$ is, in contrast, almost insensitive to the remaining small fraction ($5$ to $10~\%$) of low-$\omega$ modes ($\omega<5$) and, therefore, to the low modulus heterogeneity, $\delta G_p$. As an important consequence, we conclude that in the glass the thermal conductivity and the BP follow distinct mechanisms and are \textit{not} correlated features. This result is exact for the classical systems investigated here, where all the vibrational modes are equally excited. For more realistic cases, however, we must take into account the Bose-Einstein statistics correctly, and more involved quantum calculations are required.

{\bf Theories based on elastic heterogeneity.} We now discuss a few implications of this work. Our results support the validity of the heterogeneous elasticity theory~\cite{schirmacher_2006,schirmacher_2007,Schirmacher_2015,Schirmacher2_2015}, where elastic heterogeneities control both the BP and the glass thermal conductivity. A recent simulation study~\cite{Marruzzo_2013,Marruzzo2_2013} has tested these theoretical predictions by studying Lennard Jones glasses at different temperatures. It could be interesting to apply the theory in the case of the present system, where elastic heterogeneities can be tuned extensively. Also, in Fig.~\ref{ehthc}, we have shown an exponential relation, connecting the extent of the elastic heterogeneities both to the life-times of the acoustic-like excitations and to the thermal conductivity. In our previous work~\cite{Mizuno_2014} we also discovered a similar relation for the low-frequency transverse acoustic-like modes in the BP range. These findings deserve a more precise explanation, and should trigger additional theoretical development in the future.

{\bf The microscopic origin of heterogeneity.} In this work we have also scrutinized possible correlations between local elastic moduli and local structural quantities (including stress, density, or nature of the local order), to clarify the microscopic origin of the elastic heterogeneity. We have found that the bulk modulus heterogeneity $(\delta K)$ is related to the spatial fluctuations of volume fraction $(\delta \phi)$ and pressure $(\delta p)$. This implies that denser and more diluted system show lower and higher compressibility, respectively. On the other hand, we have not been able to highlight any effective predictor for the shear moduli heterogeneities ($\delta G_p$, $\delta G_s$). Indeed, we found some degree of correlation of the affine components with $\delta \phi$ and $\delta p$. These correlations, however, are erased by the development of the non-affine components. Identifying local quantities which are precursors of the local shear moduli heterogeneities is still an open issue~\cite{yoshimoto_2004,Tong_2014}.

{\bf Relevance for ultra-stable glasses.} The experimental work of Ref.~\cite{Swallen_2007} has demonstrated that glasses prepared by vapour deposition show extreme stability, which corresponds to equilibrium states of ordinary glasses after an aging process on time scales of thousands of years. For this reason these materials are dubbed as {\em ultra-stable} glasses~\cite{Swallen_2007,Singh_2013}. In Ref.~\cite{Singh_2013}, a numerical simulation study of ultra-stable glasses was reported, showing that the BP is reduced compared to the ordinary glasses. Differences in the local structure were also detected in the two cases. Additional work is needed to quantify in details the local elastic response in ultra-stable glasses, and highlight possible differences compared to the ordinary case.

{\bf Relevance for jammed systems.} The BP~\cite{OHern_2003,Silbert_2005}, acoustic-like excitations (Ioffe-Regel limit)~\cite{Wang2_2015}, glass-like $T$-dependence of $\kappa$~\cite{Xu_2009,Vitelli_2010}, and elastic heterogeneities~\cite{Mizuno2_2015} have been also studied in a-thermal jammed systems. As the packing fraction $\phi$ tends to the transition point, $\phi_c$, a BP progressively develops with the frequency $\Omega_\text{BP}$ vanishing~\cite{OHern_2003,Silbert_2005} (as also observed in experiments~\cite{Chen_2010}), and the transverse Ioffe-Regel frequency decreases towards zero~\cite{Wang2_2015}. Those results imply the existence of a diverging length scale~\cite{Silbert_2005,Wang2_2015}, accompanying both features. Interestingly, Refs.~\cite{Xu_2009,Vitelli_2010} have reported some degree of correlation between the BP and the $T$-dependence of $\kappa$.  Also, a recent work~\cite{Mizuno2_2015} reported that the spatial fluctuations of shear modulus diverge with vanishing global shear modulus as $\phi$ goes to $\phi_c$, which can be related to the growing BP and vanishing transverse Ioffe-Regel frequency.

In addition, in Refs.~\cite{Wyart_2005,Wyart_2006,Wyart_2010,DeGiuli_2014} a theoretical picture has been developed where the BP and glass-like thermal conductivity originate from the weak connectivities of particles (isostatic feature), due to the vicinity of the jamming transition point. We believe that a connection must exist between the elastic heterogeneities investigated here and those weak connectivities. Addressing directly this issue is an important open direction for future work.

{\bf Unified understanding of ordered and disordered solids.} We have focused on a toy model able to generate states of matter ranging from the perfect crystal, through defective crystal phases, to fully developed amorphous structures, by tuning a well designed form of particles size disorder. This choice partly follows an increasingly used methodological attitude, where data from disordered systems are systematically compared to those coming from the corresponding well-known crystalline counterparts. This approach has been employed, for instance, in the case of a-thermal jammed system in a previous work~\cite{Silbert_2006}, where the effect of structural modifications on the distribution of contact forces was systematically studied. Other recent works~\cite{Zargar_2014,Goodrich_2014,Tong_2015,Babu_2015} have followed this direction, providing a deeper understanding of important properties of materials in their crystalline and amorphous forms. Finally, in Ref.~\cite{Chumakov_2014} the vDOS and the specific heat of various glassy and crystalline polymorphs of $\text{SiO}_\text{2}$ were systematically compared. We believe that trying to connect completely ordered to disordered structures, highlighting the important variations {\em continuously}, is a fruitful line of action.

{\bf Lower-than-amorphous limit of thermal conductivity.} As a final remark, modern technologies, such as thermal management in electronic devices or thermoelectric energy conversion, employ materials with very low thermal conductivity~\cite{Venkatasubramanian_2001,Minnich_2009,Maldovan_2013}. We have demonstrated that size disorder can indeed reduce $\kappa$ towards the glass value~\cite{Mizuno2_2013}. Similar conclusions have been drawn in experimental works~\cite{Cahill_1988,Cahill_1992}, where the disorder was controlled by tuning the chemical composition of the material. In these cases, $\kappa$ is found to reach a minimum value in well-developed amorphous states, where the life-times of the heat carriers are of the order of the time scale of thermal vibrations and their mean-free-paths approach the particles sizes. 

It has been shown, however, that one can reduce the thermal conductivity even below the amorphous limit, by an appropriate design at the nano-scale of ordered systems~\cite{Hopkins_2011,Maldovan_2013}. This possibility is a crucial opportunity~\cite{Goodson_2007}, which would allow to devise (meta-)materials which are excellent thermal insulators while preserving good electronic properties, as needed in many applications~\cite{Venkatasubramanian_2001,Minnich_2009,Maldovan_2013}. Remarkably, recent experiments~\cite{Costescu_2004,Chiritescu_2007,Pernot_2010} have measured ultra-low values of $\kappa$, suggested to be smaller than the amorphous limit. These results have been confirmed by recent simulation works~\cite{Mizuno_2015,Wang_2015} demonstrating ultra-low $\kappa$ in wisely designed super-lattice nano-structures.
\begin{acknowledgments}
We acknowledge useful discussions with L.~E.~Silbert, O.~N.~Bedoya-Martinez, and A.~Onuki. This work was supported by the Nanosciences Foundation of Grenoble. J.-L. B is supported by the Institut Universitaire de France. Most of the calculations presented in this paper were performed using the Froggy platform of the CIMENT infrastructure (https://ciment.ujf-grenoble.fr), which is supported by the Rh\^one-Alpes region (GRANT CPER07 13 CIRA) and the Equip@Meso project (reference ANR-10-EQPX-29-01) of the "Programme Investissements d'Avenir", supervised by the Agence Nationale pour la Recherche.
\end{acknowledgments}
\bibliographystyle{apsrev4-1}
\bibliography{reference}
\end{document}